\documentclass[prb,reprint]{revtex4-2}
\usepackage{graphicx}
\usepackage{enumitem}
\usepackage{amsmath}
\newcommand{\ket}[1]{|{#1}\rangle}
\newcommand{\bra}[1]{\langle{#1}|}
\newcommand{\braket}[1]{\langle{#1}}
\usepackage{amsfonts}

\usepackage[usenames,dvipsnames]{xcolor}
\usepackage[colorlinks=true,citecolor=Cerulean,linkcolor=RubineRed,urlcolor=Cerulean]{hyperref}

\begin{document}

\title{Counterdiabatic-influenced Floquet-engineering: \\ State preparation, annealing and learning the adiabatic gauge potential}

\author{Callum W. Duncan}
\email{callum.duncan@aegiq.com}
\affiliation{Department of Physics and SUPA, University of Strathclyde, Glasgow G4 0NG, United Kingdom}
\affiliation{Aegiq Ltd., Cooper Buildings, Arundel Street, Sheffield, S1 2NS, United Kingdom}

\begin{abstract}
Counterdiabatic driving, which enforces adiabatic evolution in arbitrary timescales, can be realised by engineering a Floquet Hamiltonian which oscillates between the Hamiltonian and its derivative requiring no additional control terms. However, the coefficients of the Floquet Hamiltoinan require knowledge of the counterdiabatic terms, which can be difficult to derive outside of a limited set of examples. We introduce a new hybrid technique for the control of quantum systems, Counterdiabatic-influenced Floquet-engineering or CAFFEINE for short. CAFFEINE parameterises the Floquet Hamiltonian for counterdiabatic driving and utilises numerical quantum optimal control in order to obtain the desired target state. This removes the need to both obtain and implement counterdiabatic terms, however, it does require the ability to quickly oscillate each term in the Hamiltonian. If this oscillation is possible, then CAFFEINE provides a framework to implement quantum annealing protocols and general quantum state preparation. We illustrate this through optimisation of two numerical examples of preparing a Bell state with two qubits and performing annealing protocols for the one-dimensional Ising model. Beyond this, we also illustrate CAFFEINE's capabilities to learn the counterdiabatic terms, which can potentially be used as a probe of quantum chaos and the geometry of quantum dynamics.
\end{abstract}
\pacs{}

\maketitle

\section{Introduction}

The ability to manipulate quantum systems to a desired target state is crucial for the implementation of quantum computing and simulation protocols on quantum hardware. The adiabatic approximation is particularly useful in this task and is a building block of both adiabatic quantum computing \cite{aharonov2008adiabatic,Ablash2018adiabatic} and quantum annealing \cite{morita2008mathematical,yarkoni2022quantum,rajak2023quantum}. It states that if a dynamical parameter change is slow then the dynamical solution to the Schr\"{o}dinger equation is given by the decomposition of the instantaneous eigenstates up to a dynamical phase factor \cite{born1928beweis,kato1950adiabatic}. This means that if we initiate a system in an eigenstate of an initial Hamiltonian then it will follow the path corresponding to that eigenstate throughout the adiabatic evolution. However, this relies upon the slow variation of the dynamical parameters, such that the gap between all states, and specifically the state of interest being followed, is large with respect to the inverse of the time taken to traverse the energy landscape \cite{jansen2007bounds}. If the energy gaps present are exponentially small, as can be the case in non-trivial many-body systems, then the adiabatic approximation requires exponential time to traverse the path.

There are many approaches to speed up adiabatic protocols, referred to collectively as shortcuts to adiabaticity \cite{gueryodelin2019shortcuts,torrontegui2013shortcuts}, as well as protocols that abandon the need for adiabatic dynamics if the sole aim is to prepare a specific target state, which forms part of the family of techniques of quantum optimal control \cite{glaser2015training,ansel2024introduction}. Counterdiabatic driving (CD) \cite{demirplak2003adiabatic,demirplak2005assisted,berry2009transitionless} is a promising technique from the field of shortcuts to adiabaticity, which adds control terms to the dynamical Hamiltonian to enforce the adiabatic approximation to be the solution of the Schr\"odinger equation in arbitrary timescales. However, exactly deriving the CD terms is difficult as it relies on knowledge of the instantaneous eigenstates across the path. As a result the exact CD is only known in a limited set of scenarios, e.g., harmonic oscillators \cite{campo2013shortcuts}, transverse Ising model \cite{campo2012assisted,damski2014counterdiabatic}, and non-interacting tight-binding models \cite{duncan2024exact}. Methods have been introduced to numerically obtain the CD terms \cite{lawrence2024numerical,Takahashi2024shortcuts}, however, even if we obtain them, these terms can require the control of interactions across distant parts of the system including up to general $N$-body interactions. This makes CD difficult to implement in all but the simplest of quantum systems. The CD terms are also equivalent to the  adiabatic gauge potential (AGP) \cite{sels2017minimizing,kolodrubetz2017geometry} which characterises all diabatic terms of a dynamical path, i.e., CD is the countering of the AGP. We note here that recently there has been growing interest in gate based quantum algorithms inspired by CD \cite{Vreumingen2024gate,li2024quantum,Hegade2022digitized}, specifically enhancements to the quantum approximate optimisation algorithm \cite{wurtz2022counterdiabaticity,Chandarana2022digitized,blekos2024review} and feedback-based quantum algorithms \cite{malla2024feedback}.

A possible approach to overcome the non-local nature of CD is to use the approach of variational, or local, CD which truncates the CD to include only low-order terms \cite{sels2017minimizing,Saberi2014adiabatic}. However, this is of course an approximate method with no guarantee that all diabatic terms are corrected. Recently, hybrid techniques combining local CD and quantum optimal control have been proposed and demonstrated some success in state preparation in large quantum many-body models which are still accessible to direct simulation \cite{cepaite2023counterdiabatic,morawetz2024efficient}. However, these hybrid approaches still rely on both the derivation and implementation of additional terms to be added to the Hamiltonian. 

In this work, we propose a hybrid approach which avoids both the need to derive and implement additional terms, relying only on control of terms in the dynamical Hamiltonian of the original problem. This is the method of CounterdiAbatic-inFluenced FloquEt-engINEering or CAFFEINE for short. This approach builds on recent advances that demonstrated that it was possible to emulate the introduction of CD via high-frequency oscillations such that the overall Floquet Hamiltonian mimics the CD Hamiltonian \cite{petiziol2018fast,petiziol2019accelerating,Villazon2019swift}. This includes a general strategy for designing fast adiabatic protocols via Floquet engineering with control of only the terms contained within the Hamiltonian \cite{claeys2019floquet}. These counderdiabatic Floquet schemes have been realised experimentally in small scale quantum hardware \cite{Boyers2019floquet,zhou2019floquet}. However, knowledge of the coefficients of the general counterdiabatic Floquet Hamiltonian obtained in Ref.~\cite{claeys2019floquet} require deriving exact or approximate CD terms. CAFFEINE instead is a hybrid method which parameterises the general counterdiabatic Floquet Hamiltonian and utilises methods of numerical quantum optimal control to obtain optimised protocols for the preparation of states and the realisation of quantum annealing.

CAFFEINE also provides a route to obtaining the AGP, and all CD terms, directly from the quantum hardware. Of course, in doing so, CAFFEINE will prepare the target state meaning that the obtained CD terms will not be required to be implemented. However, this could provide a useful novel probe of quantum systems, as the AGP has been shown to be related to properties of chaotic quantum systems \cite{Pandey2020adiabatic,lim2024defining}, to be sensitive to the presence of a quantum phase transition \cite{lawrence2024numerical,Hatomura2021controlling}, to probe the closing of band gaps \cite{duncan2024exact,balducci2024fighting}, and to generally study the geometry of nonequilibrium behaviour \cite{kolodrubetz2017geometry}. 

In this work we will begin by expanding upon the approach of CD and the AGP before introducing CAFFEINE in Sec.~\ref{sec:CAFFEINE}. We then consider in detail a two qubit example where we prepare Bell states and consider CAFFEINE's ability to learn the AGP. We finish in Sec.~\ref{sec:Ising} by considering modest size 1D nearest neighbour Ising models, showing that for increasing system size we can scale up CAFFEINE in order to obtain increasing improvements in the final result.

\begin{figure*}[ht]
	\centering
	\includegraphics[width=0.98\linewidth]{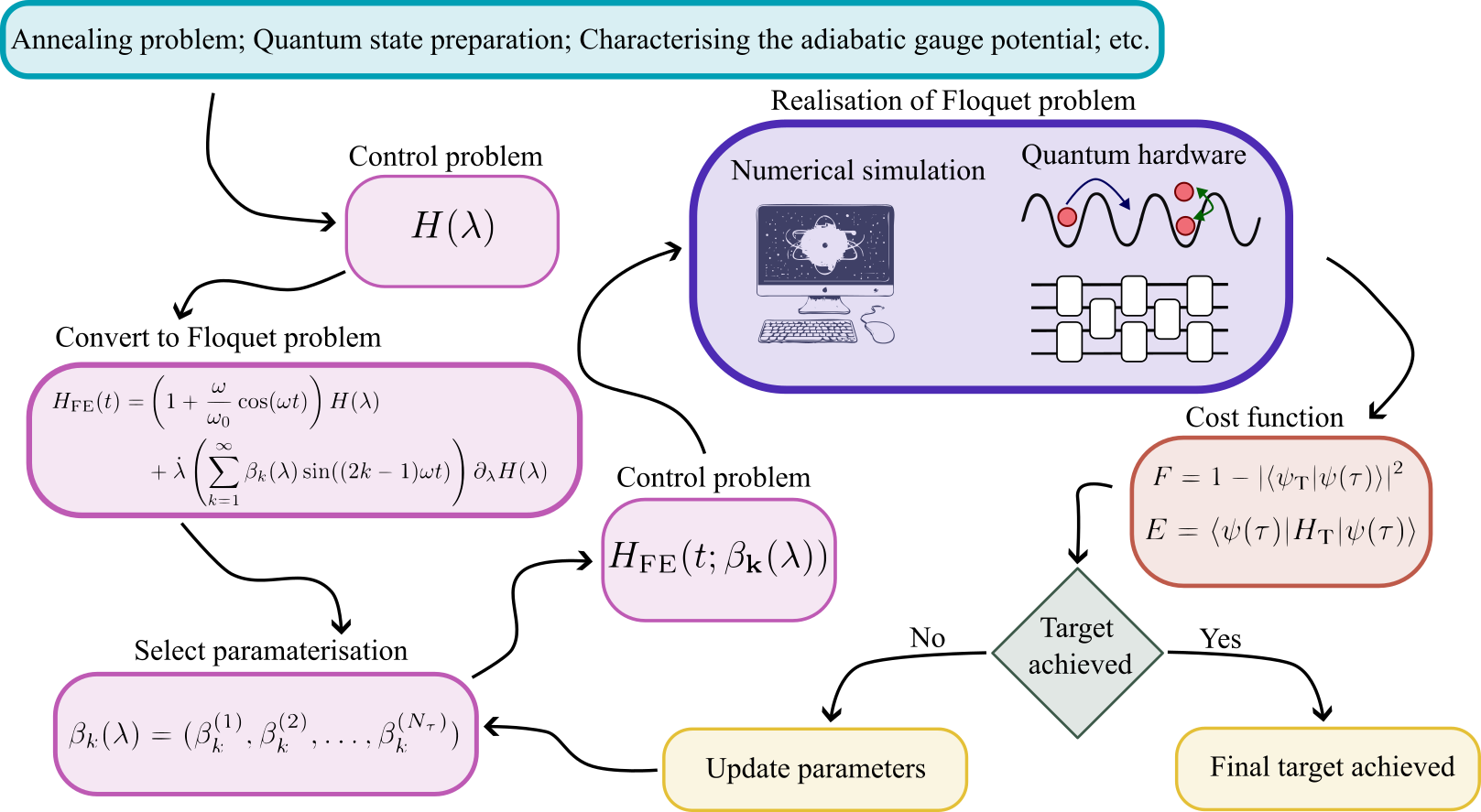}
	\caption{Illustration of the workflow of the CounterdiAbatic-inFluenced FloquEt-engINEering or CAFFEINE approach. While we will only numerically simulate the quantum evolution of the periodically-driven system, we envision that CAFFEINE could be implemented as a closed-loop protocol with quantum hardware.}
	\label{fig:illustration}
\end{figure*}

\section{Counterdiabatic driving and the adiabatic gauge potential}

We will consider the general case of a Hamiltonian $H(\lambda)$ dependent on a parameter, or family of parameters, $\lambda$ where we aim to start in an initial eigenstate of $H(\lambda_i)$ and reach a final eigenstate of $H(\lambda_f)$. This is the case for standard quantum annealing problems, which we will consider below, but also for general state preparation. The most conceptually straight forward approach to this problem is to consider $\lambda(t)$ varying from $\lambda_i$ to $\lambda_f$ on an adiabatic timescale. However, this approach is not possible in many settings primarily due to the exponentially small gaps of quantum many-body systems requiring exponentially long protocols to traverse adiabatically.

One approach to tackle the issue of diabatic excitations is CD \cite{demirplak2003adiabatic,demirplak2005assisted,berry2009transitionless}, which can be realised by adding a particular auxilary control term resulting in the new Hamiltonian
\begin{equation}\label{eq:HCD}
    H_\mathrm{CD}(\lambda) = H(\lambda) + \dot{\lambda} \mathcal{A}_\lambda \ ,
\end{equation}
where $\mathcal{A}_\lambda$ is the AGP. The AGP exactly describes all diabatic transitions between eigenstates and can be defined in terms of the instantaneous eigenstates and eigenenergies obtained from the time-independent Schr\"odinger equation $H(\lambda) \ket{n} = E_n \ket{n}$ as
\begin{equation} \label{eq:AGP}
    \bra{m} \mathcal{A}_\lambda \ket{n} = i \braket{m} \ket{\partial_\lambda n} = -i \frac{\bra{m} \partial_\lambda \ket{n}}{E_m - E_n} \ ,
\end{equation}
with $\partial_\lambda$ denoting the partial derivate with respect to $\lambda$. The central issues with CD is clear from Eq.~\eqref{eq:AGP}. First, calculating the AGP from Eq.~\eqref{eq:AGP} requires knowledge of not only the eigenspectrum but the eigenstates across the full time-dependent protocol parameter $\lambda$. Second, even if this is possible, which it is not in general, the difference in energy levels can become exponentially small in scenarios where adiabatic dynamics is difficult, e.g. across phase transitions, leading to divergent terms in the AGP. As a result, it is common for the AGP of quantum many-body systems to contain highly non-local terms \cite{sels2017minimizing,lawrence2024numerical} which are difficult to control in current, and likely future, experiments.

As described in the introduction, there has been recent key developments in our understanding of the AGP and our ability to calculate it numerically. For the most part this is due to the realisation that the AGP can be defined through a general ansatz of \cite{claeys2019floquet} 
\begin{equation}\label{eq:commansatz}
    \mathcal{A}_\lambda = i\sum_{k=1}^\infty \alpha_k \underbrace{\Big[H(\lambda),\big[H(\lambda),\ldots[H(\lambda)}_{2k-1},\partial_\lambda H(\lambda)]\big]\Big] \ ,
\end{equation}
where $\alpha_k \in \mathbb{C}$ is a set of coefficients which we still need to solve for. This commutator ansatz was immediately combined with the variational approach of local CD to determine the coefficients up to a given cutoff in $k$ \cite{claeys2019floquet}. Since then, it has also resulted in new algorithms for calculating the exact AGP \cite{lawrence2024numerical} and for a Krylov space approach \cite{Takahashi2024shortcuts,morawetz2024efficient}. However, this does not address the issue that it is in general hard to implement the exact AGP to realise counterdiabatic driving in quantum many-body systems.

It has been found that it is possible to implement the counterdiabatic Hamiltonian of Eq.~\eqref{eq:HCD} with the exact AGP via a Floquet engineered Hamiltonian of the form \cite{claeys2019floquet}
\begin{equation}\label{eq:HFE}
\begin{aligned}
    H_\mathrm{FE} (t)  = & \left( 1 + \frac{\omega}{\omega_0} \cos (\omega t) \right) H(\lambda) \\ & + \dot{\lambda} \left( \sum_{k=1}^{N_k} \beta_k(\lambda) \sin((2k-1) \omega t) \right) \partial_\lambda H(\lambda) \ ,
\end{aligned}
\end{equation}
where $\omega_0$ is a reference frequency and $\beta_k(\lambda)$ are Fourier coefficeints of the additional drive. In the infinite-frequency limit $\omega\rightarrow \infty$, $H_\mathrm{FE}$ with $N_k \rightarrow \infty$ realises stroboscopic counterdiabatic driving as was shown in Ref.~\cite{claeys2019floquet}. Note, in this case the time scale characterising the oscillations $2\pi/\omega$ is far faster than the timescale of the protocol of driving $\lambda$ resulting in a decoupling of the dynamics. By following some algebraic steps it can be seen that the $\beta_k(\lambda)$ coefficients is proportional to the $\alpha_k$ coefficients of the AGP. The proposal was then to derive the coefficients of the AGP and to use these to find the Fourier coefficients for the Floquet engineered Hamiltonian of Eq.~\eqref{eq:HFE}. Note, for any finite size system we will be able to take a finite cutoff of $N_k$ and retain the realisation of the exact AGP in that system, or we can take a lower cutoff and realise a similar protocol to that of the approximate local CD \cite{sels2017minimizing}.

\section{CAFFEINE: Counterdiabatic-influenced Floquet-engineering}\label{sec:CAFFEINE}

CounterdiAbatic-inFluenced FloquEt-engINEering (CAFFEINE) takes the step to remove the requirement that we need to solve for or implement the AGP. Instead, we will use the Floquet engineered Hamiltonian of Eq.~\eqref{eq:HFE} and treat $\beta_k(\lambda)$ as a control parameter to be optimised for the desired outcome. The general approach of CAFFEINE is outlined in this section and illustrated in Fig.~\ref{fig:illustration}. For state preparation, this then becomes simply an optimal control problem for the $\beta_k(\lambda)$ up to a considered cut-off. Note, CAFFEINE for state preparation could be seen as a simplification down to specific terms which are known to induce the counterdiabatic terms compared to the more general fast adiabatic evolution by oscillation described in Ref.~\cite{petiziol2018fast}.

The choice of parametrisation of $\beta_k(\lambda)$ is arbitrary, and can be tailored to fit particular problem instances. We will consider taking $\beta_k(\lambda)$ to be a piecewise-constant function, 
\begin{equation}
\beta_k(\lambda) = (\beta_k^{(1)},\beta_k^{(2)},\ldots,\beta_k^{(N_\tau)}) \ ,
\end{equation}
where $N_\tau$ is the total number of time steps of $\beta_k(\lambda)$ across $\lambda$ and
\begin{equation}\label{eq:betadiscrete}
\beta_k(\lambda) = \beta_k^{(j)} \ \text{if}\ (j-1)\Delta \lambda \leq \lambda < j\Delta \lambda \ ,
\end{equation}
where $\Delta \lambda$ is the parameter step, which we will consider to be constant and given by $\Delta \lambda = \tau/N_\tau$ with $\tau$ the total protocol time. CAFFEINE is compatible with more advanced quantum optimal control approaches, e.g., CRAB \cite{Caneva2011chopped,Rach2015dressing,muller2022one}, GRAPE \cite{khaneja2005optimal,Motzoi2011optimal,lu2024optimal} or reinforcement learning \cite{bukov2018reinforcement,niu2019universal,zhang2019does,sivak2022model}, which we will not explore here.

CAFFEINE opens a different intriguing possibility in addition to a new approach to state preparation and the realisation of annealing protocols. While there have been significant algorithmic improvements in obtaining the AGP numerically with the study of Krylov \cite{Takahashi2024shortcuts} and other methods \cite{lawrence2024numerical}, these will all still be fundamentally limited by what is possible to solve and simulate using a conventional `classical' computer. With CAFFEINE we can propose a learning-style approach to obtaining the AGP. This can be performed by realising the CAFFEINE protocol with quantum hardware then optimising for $\beta_k(\lambda)$ in the same manner as a variational quantum algorithm. After obtaining the $\beta_k(\lambda)$ these can be related to the $\alpha_k$ of the AGP given in Eq.~\eqref{eq:AGP} in an inverted approach to that outlined in Ref.~\cite{claeys2019floquet}. Essentially, CAFFEINE for the AGP utilises a discretised cost function which attempts to, e.g., minimise the energy across a number of segments of the dynamics, i.e., the cost function is written as a function of each $\beta_k^{(j)}$ as defined in Eq.~\eqref{eq:betadiscrete} instead of the full $\beta_k(\lambda)$. We will consider an example and the protocol for learning the AGP in detail in Sec.~\ref{sec:AGPlearn}.

\section{Examples}

We will consider the general class of problems described by the standard quantum annealing Hamiltonian
\begin{equation}\label{eq:anneal}
    H(\lambda(t)) = (1-\lambda(t)) H_m + \lambda(t) H_p
\end{equation}
with $\lambda(t)$ the single parameter being varied in time with $\lambda(0) = 0$ and $\lambda(\tau) = 1$ where $\tau$ is the total time of the protocol. Note, we can consider both state preparation and annealing problems with this Hamiltonian and CAFFEINE is generalisable to Hamiltonians with a family of parameters changing in time. 

Numerical calculation of the quantum dynamics will be performed using the QuTiP python library \cite{johansson2012qutip} and we will in general utilise the SciPy dual annealing algorithm \cite{2020SciPy-NMeth} for the optimisation algorithm with a maximum number of function evaluations of $10^5$ and global search iterations of $10^{3}$. We note, that considering the large frequency limit of Floquet engineering dynamically is in general computationally intensive and the results presented here are mainly limited by the time required to calculate a single iteration of the optimisation algorithm.

\subsection{Two-qubit entanglement generation}

As a first example we will consider the two qubit system given by the Hamiltonian 
\begin{equation}\label{eq:2qubit}
    H(\lambda) = -J (\sigma^x_1 \sigma^x_2 + \sigma^z_1 \sigma^z_2) + h_z (\lambda - 1) (\sigma^z_1 + \sigma^z_2) \ ,
\end{equation}
with $J$ the coupling strength, $h_z$ the z-field, and $\sigma^i$ the Pauli matrices of the $i$th type. We will consider units of $J$ and take $h_z=5J$ and consider protocols of length $\tau=0.1J^{-1}$ with the parameter $\lambda$ given by
\begin{equation}\label{eq:lambdasin}
    \lambda(t) = \sin^2 \left( \frac{\pi}{2} \sin^2 \left( \frac{\pi t}{2 \tau}\right)\right) \ ,
\end{equation}
which satisfies $\lambda(0)=0$ and $\lambda(\tau)=1$ with the time derivatives $\dot{\lambda}(0) = \dot{\lambda}(\tau) = 0$, which generally is helpful for implementing adiabatic pulses. We select this particular 2-qubit example as it has been considered by previous protocols, in part as we can analytically obtain the adiabatic gauge potential as \cite{claeys2019floquet,petiziol2019accelerating}
\begin{equation}
    \mathcal{A}_\lambda = \frac{J h_z}{2\left( J^2 + 4(\lambda-1)^2 h_z^2 \right)} \left( \sigma^y_1 \sigma^z_2 + \sigma^z_1 \sigma^y_2 \right) \ .
\end{equation}
This allows one to construct the exact Floquet-engineered Hamiltonian for counterdiabatic driving \cite{claeys2019floquet} by taking Eq.~\eqref{eq:HFE} with $\beta_{k\neq 1}(\lambda)=0$ and
\begin{equation}\label{eq:analybeta}
    \beta_1(\lambda) = \frac{2 h_z \omega_0}{4J^2 + 16(\lambda(t)-1)^2 h_z^2} \ .
\end{equation}
For this example, for CAFFEINE we will focus on the optimisation of $\beta_1(\lambda)$ as from this analytical form we can set higher order terms to zero. Note, that this is not always the case for 2-qubit Hamiltonians, with the transverse Ising model for 2-spins being described by $\beta_1 (\lambda) \neq 0$ and $\beta_2 (\lambda) \neq 0$. For more general many-body Hamiltonians, it is normally the case that many of the $\beta_k(\lambda)$ coefficients will be non-zero, as the AGP can require the support of many terms in the commutator expansion \cite{lawrence2024numerical}.

\subsubsection{State preparation}\label{sec:stateprep}

We will first consider state preparation for the 2-qubit Hamiltonian given by Eq.~\eqref{eq:2qubit}, for which we start at $t=0$ in the trivial state $\ket{\uparrow \uparrow}$ and end at $t=\tau$ at the Bell state $(\ket{\uparrow \uparrow} + \ket{\downarrow \downarrow})/\sqrt{2}$, if the protocol is performed perfectly adiabatically. While this might be a tractable problem, it can form a building block of other more complex Hamiltonians and dynamics, e.g., entanglement generation. 

\begin{figure*}[ht!]
	\centering
	\includegraphics[width=0.8\linewidth]{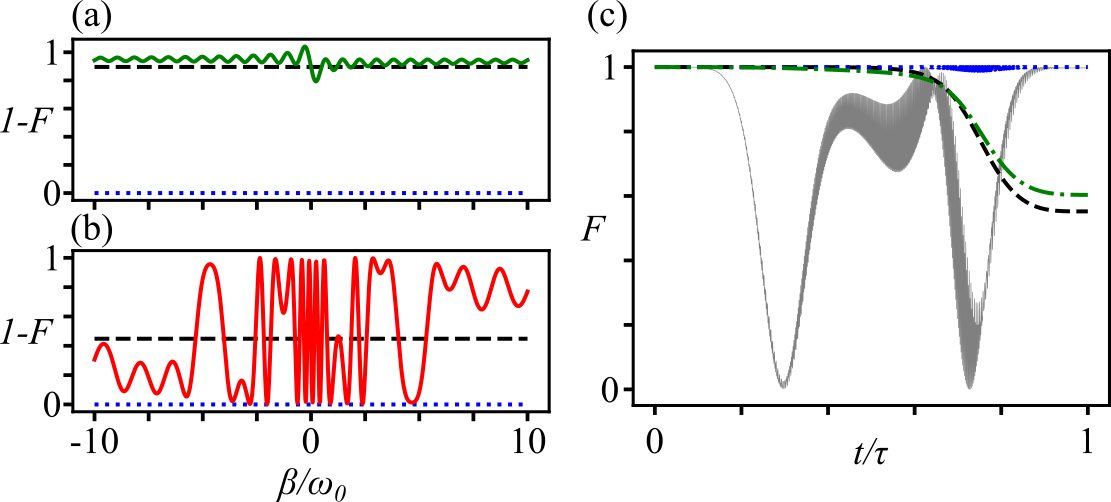}
	\caption{Two-qubit entanglement generation cost function landscape and optimal solution for $N_\tau=1$ parameters in time. (a) the cost function landscape for the optimised annealing shown by solid (green) lines with the analytical Floquet engineered shown by a dotted (blue) line and the unassisted annealing shown by a dashed (black) line. (b) The same as (a) but with the optimised Counterdiabatic-inspired Floquet-engineering (CAFFEINE) shown with a solid (red) line. (c) Fidelity with respect to the instantaneous eigenstate at time $t$ for the case of analytical Floquet-engineering shown by a dotted (blue) line, the bare annealing protocol with a dashed (black) line, the optimised annealing with a dash-dot (green) line, and the optimised CAFFEINE by the grey region. The values of $\gamma_1$ and $\beta_1$ for the optimised annealing and CAFFEINE are taken from the minimum values found in the sweep of the cost function landscape shown in (a) and (b) respectively.}
	\label{fig:Fig1}
\end{figure*}

\begin{figure}[ht!]
	\centering
	\includegraphics[width=0.8\linewidth]{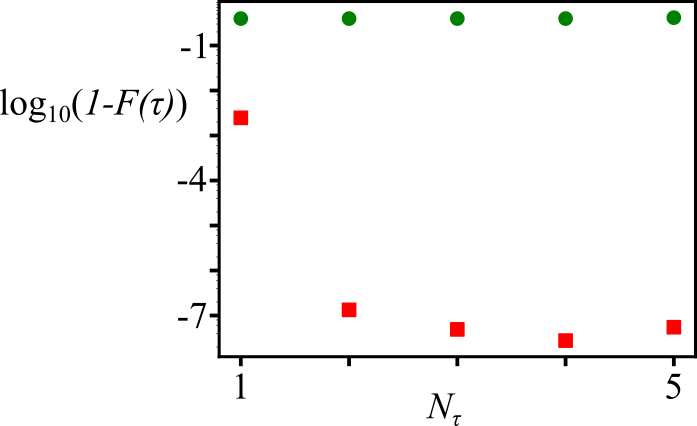}
	\caption{Two-qubit entanglement generation with increasing number of timesteps $N_\tau$ for the piecewise-constant coefficient $\beta_1$. The optimised CAFFEINE results are shown by squares (red) with the optimised annealling protocol including the local Z-fields of Eq.~\eqref{eq:control} are shown by circles (green).}
	\label{fig:Fig2}
\end{figure}

For this example of state preparation we will consider a comparison of CAFFEINE to optimised annealing by adding a control term to Eq.~\eqref{eq:2qubit} of the form
\begin{equation}\label{eq:control}
    H_c(t) = \sum_{k=1}^{N_\tau} \gamma_k \sin\left( \frac{2 \pi k t}{\tau} \right) \left( \sigma^z_1 + \sigma^z_2 \right) \ ,
\end{equation}
with $\gamma_k$ the coefficients optimised over. In selecting this control function, we do not intend to consider the best optimal control protocol without Floquet-engineering, but instead intend to inform whether the additional controls of the CAFFEINE Hamiltonian of Eq.~\eqref{eq:HFE}, i.e. local z-fields, are playing a significant role or if it is truly the counterdiabatic terms introduced by the Floquet-engineering. For the latter, we would expect the optimised annealing with the control term of Eq.~\eqref{eq:control} to have far less of an impact on the fidelity than that of CAFFEINE, which we will show is indeed the case in this section.

We will first consider the general form of the cost function landscape for CAFFEINE optimisation problems by considering a single timestep, i.e. $N_\tau = 1$, for $\beta_1$, meaning $\beta_1$ is a constant and not $\lambda$-dependent. Our goal here is to prepare the state with a high fidelity, $F = |\braket{\psi_T} \ket{\psi(\tau)}|^2$ with $\ket{\psi_T}$ the target state (the 2-qubit Bell state $(\ket{\uparrow \uparrow} + \ket{\downarrow \downarrow]})/\sqrt{2}$) and $\ket{\psi(\tau)}$ the final state after the dynamical protocol at time $\tau$. To achieve this we do not require that the dynamics are adiabatic throughout the protocol, simply that we prepare the state at the end, hence we will use the fideilty with respect to the target state as the cost function.

The cost function landscape for the optimised annealing is shown in Fig.~\ref{fig:Fig1}a. We see from this that a single control term for the optimised annealing has little impact on the fidelity, but in the range of parameters there is a clear global minimum which achieves a slight improvement in fidelity. However, there are still many local minima in the landscape and it is non-convex. At its optimal value in the probed landscape, $\gamma_1=0.22\omega_0$, optimised annealing achieves a state preparation fidelity of $1-F=0.397$, which is marginally better than unassisted annealing, i.e. $\gamma_1=0$, which achieves $1-F=0.448$. Note, that both of these are consistent with the dynamical state for the most part still having significant overlap with the initial state. 

For CAFFEINE it is immediately clear that the landscape, shown in Fig.~\ref{fig:Fig1}b, is more complicated, and that the fidelity can vary over small parameter regimes between the most extreme values possible. It is clear though that CAFFEINE can achieve extremely good fidelity even when restricted to one single value of the optimisation parameters as a function of time. The optimal value probed for CAFFEINE in this landscape, $\beta_1=-2.02\omega_0$, results in a final state preparation fidelity of $1-F=2.47\times10^{-3}$ with the analytical Floquet-engineering with the time-dependent $\beta_1(\lambda)$ given by Eq.~\eqref{eq:analybeta} achieving $1-F=5.87\times10^{-6}$. While there is quite some distance between the fidelities of CAFFEINE and the analytical exact approach, this is a first demonstration of the power of CAFFEINE alone, as without even implementing a time dependent $\beta_k$ paramaeter, we can achieve Bell state preparation to approximately $F=0.997$.

The fidelity with respect to the instantaneous eigenstate at time $t$ for each protocol is shown in Fig.~\ref{fig:Fig1}c. The different game being played by CAFFEINE is then clear, with the instantaneous fidelity getting as low as $F=8.51\times10^{-6}$ as the dynamical state takes a very different path through the family of dynamical Hamiltonians that are now available. This is in stark contrast to the analytical Floquet-engineered pulse, which is intended to keep the dynamics adiabatic throughout the protocol.

We now consider increasing the number of possible values during the protocol of the piecewise-constant function $\beta_1$ as described by Eq.~\eqref{eq:betadiscrete}. The achieved fidelity for both the optimised annealing and CAFFEINE are shown in Fig.~\ref{fig:Fig2}. The optimised annealing does not provide a noticeable improvement to the state preparation protocol for any tested value of $N_\tau$. This is evidence that any improvement given by the CAFFEINE protocol is not due to the additional control of local z-fields in this example, but due to the additional terms that are generated from the high-frequency dynamics. At the same time, introducing the most simple of time-dependence into the optimised $\beta_1$ of CAFFEINE results in Bell state preparation to approximately $F=0.9999999$, or $1-F=10^{-7}$.

\subsubsection{Learning the adiabatic gauge potential}
\label{sec:AGPlearn}

We now turn to another possible application of CAFFEINE, in learning the AGP in the first place. As discussed in Sec.~\ref{sec:CAFFEINE} and Ref.~\cite{claeys2019floquet}, the $\beta_k(\lambda)$ fully characterise the AGP. Therefore, if we can learn the optimal $\beta_k(\lambda)$ protocols for a given $H(\lambda)$ then we will have obtained the AGP for that $H(\lambda)$. This is a paradigm shift compared to the usual approaches of obtaining the AGP, where we can learn the AGP from the dynamics of quantum hardware. We also note, that this is obtained while also potentially solving the annealing problem of the same system, which we will detail below.

We will consider the 2-spin Hamiltonian of Eq~\eqref{eq:2qubit}, as we can analytically obtain $\beta_1$, given in Eq.~\eqref{eq:analybeta}, for comparison. We note that the true power of this approach will come in examples where the exact AGP is not analytically known. To learn the AGP we implement CAFFEINE as before but perform the annealing protocol in series for each of the $N_\tau$ steps of $\beta_1(\lambda)$. Thereby enforcing approximately adiabaitic dynamics throughout the applied drive. This can be easily generalised to higher order terms, i.e. $\beta_{k\neq1}(\lambda)$. By perfoming the annealing protocol in each step by minimising at the $j$th step $E_j=\bra{\psi((j+1)\Delta\lambda)} H((j+1)\Delta\lambda) \ket{\psi((j+1)\Delta\lambda)}$ where $\ket{\psi((j+1)\Delta\lambda)}$ is the dynamical state at the end of the $j$th protocol between $t_\mathrm{s}=j\Delta\lambda$ and $t_\mathrm{f}=(j+1)\Delta\lambda$ where the intial state was $\ket{\psi(j\Delta\lambda)}$. During each optimisation step we will consider $0\leq\beta\leq\omega_0$.

\begin{figure}[t!]
	\centering
	\includegraphics[width=0.98\linewidth]{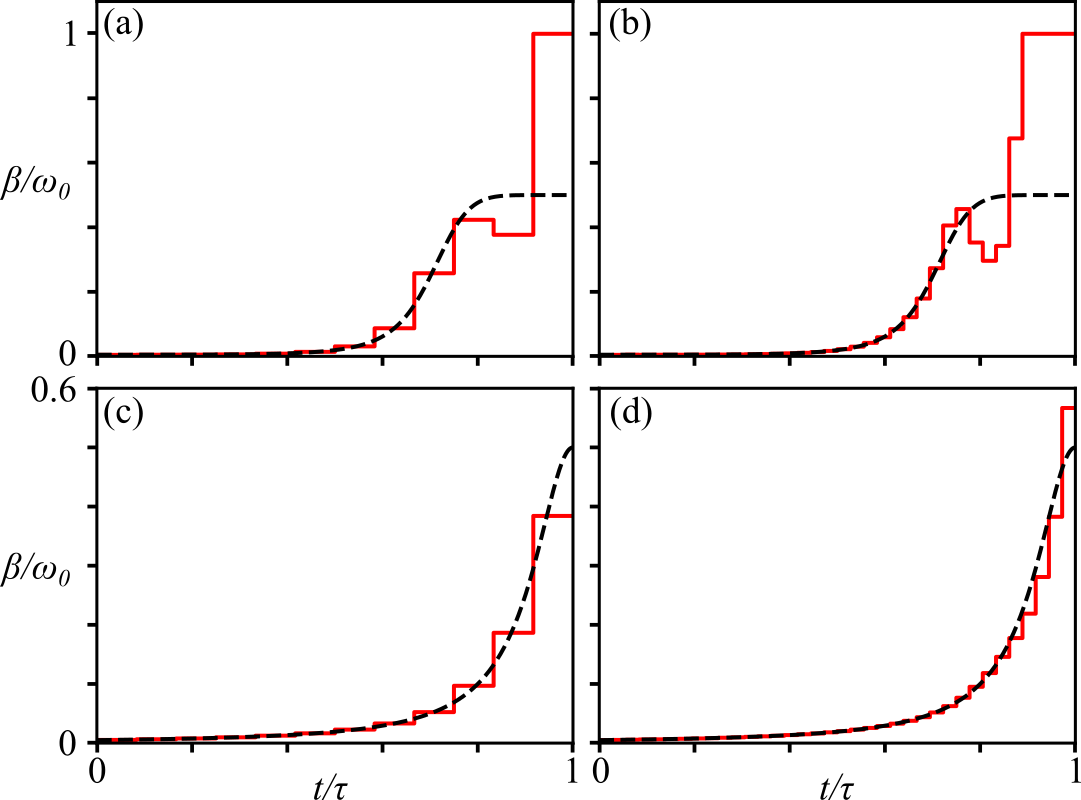}
	\caption{Learning the adiabatic gauge potential from an annealing schedule for a two-qubit dynamical Hamiltonian. Analytically derived values of $\beta$, which fully characterises the adiabatic gauge potential, are shown by dashed (black) lines with the learned $\beta$ shown by solid (red) lines. (a) Shows the case of $N_\tau=12$ for the smooth protocol, (b) the same as (a) but with $N_\tau=36$, (c) the case of $N_\tau=12$ for the linear protocol, and (d) the same as (c) but with $N_\tau=36$}
	\label{fig:Fig4}
\end{figure}

First we consider the same protocol as was considered for the state preparation in Sec.~\ref{sec:stateprep}, i.e. $\lambda$ is given by Eq.~\eqref{eq:lambdasin}. The comparison between the analytical and learned $\beta$ are shown in Fig.~\ref{fig:Fig4}a and b for $N_\tau=12$ and $36$ respectively. We observe that for $t/\tau \lesssim 0.8$ that we achieve good agreement between the piecewise-constant approximation which has been learned and the analytical $\beta_1(\lambda)$. However, at the final region of the protocol, $t/\tau \gtrsim 0.8$, we see large deviations between the analytical and learned. The choice of restricting $0\leq\beta\leq\omega_0$ was entirely motivated so as to stop the fast growth of the value of the piecewise constant at the end of the drive. While at first worrying, these deviations are in fact entirely caused by the choice of $\lambda(t)$ in Eq.~\eqref{eq:lambdasin} for which we recall that the time derivatives $\dot{\lambda}(0) = \dot{\lambda}(\tau) = 0$ and they will continuously tend towards these values at the start and end of the protocol. Since the dynamical protocol given by Eq.~\eqref{eq:HFE} is truly implementing the product $\dot{\lambda} \beta_k(\lambda)$, this means that while the analytical $\beta_k(\lambda)$ is non-zero, the implemented term is tending towards zero at the end of the protocol. As a result, the learned $\beta_1(j\Delta \lambda)$ for $j\Delta\lambda \gtrsim 0.8$ are irrelevant and we should not expect CAFFEINE to be able to learn the values of the AGP in this region.

As a comparison, we consider a linear pulse for the protocol with
\begin{equation}
    \lambda_\mathrm{lin}(t) = \frac{t}{\tau} \ ,
\end{equation}
which has a constant time derivative of $\dot{\lambda}_\mathrm{lin} = 1/\tau$ and should not encounter the same problems as the choice of Eq.~\eqref{eq:lambdasin}. This is indeed observed in Fig.~\ref{fig:Fig4}c and d, with CAFFEINE able to learn the AGP in good agreement with the expected analytical form.

\subsection{Ising model}\label{sec:Ising}

For many annealing protocols, the solution is encoded in the ground state of an Ising Hamiltonian \cite{Kadowaki1998quantum,lucas2014ising,rajak2023quantum}, this includes the study of QUBO problems. Here, we will consider as an application of CAFFEINE annealing problems of the form of Eq.~\eqref{eq:anneal} with
\begin{equation}
    H_m = - \sum_i \sigma_i^x \ ,
\end{equation}
and
\begin{equation}
    H_p = - \sum_{i<j} J_{ij} \sigma_i^z \sigma_j^z + \sum_i^N h_i \sigma_i^z \ ,
\end{equation}
where $J_{ij}$ and $h_i$ are the coupling and local fields respectively. While it is possible to obtain low-order approximations to the AGP for this form of Hamiltonian, it has been shown that for the general spin-glass Hamiltonians with arbitrary connectivity, i.e., $J_{ij}$ drawn from a random distribution of size $N$, the supporting operator-space of the AGP grows exponentialy with $N$ \cite{lawrence2024numerical}. We will consider the case of the 1D nearest neighbour Ising model with $j=i+1$ and leave the investigation of other models to future studies. We will take units of energy $J$ with $\lambda$ to be given by Eq.~\eqref{eq:lambdasin} with $\tau=0.1J^{-1}$, all $h=0$. We will consider the minimisation of the energy of the final prepared state as is the usual practice for quantum annealing.

\begin{figure}[t]
	\centering
	\includegraphics[width=0.98\linewidth]{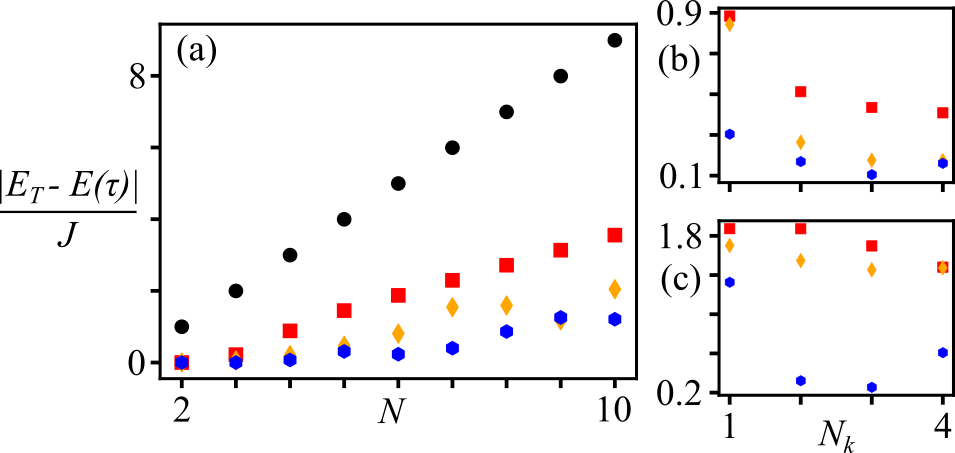}
	\caption{Quantum annealing in the Ising model. (a) Energy of final state compared to the the ground state energy, $E_T$, for various system sizes $N$ after the implementation of CAFFEINE. Circles (black) show the unassisted case for comparison to CAFFEINE with only $N_k=1$ and square (red) points giving $N_\tau = 1$, diamond (orange) giving $N_\tau = 6$, and hexagon (blue) giving $N_\tau = 12$ timesteps. We explore increasing the frequency of the oscillating term by increasing $N_k$ for fixed (b) $N=4$ and (c) $N=6$, with square (red) points giving $N_\tau = 1$, diamond (orange) giving $N_\tau = 2$, and hexagon (blue) giving $N_\tau = 3$ timesteps.}
	\label{fig:Fig5}
\end{figure}

We note here that while the exploration of the 1D nearest neighbour Ising model for up to $N=10$ as conducted here is rather modest, the simulation of the time-dependent and quickly oscillating Hamiltonian for the case of CAFFEINE, given in Eq.~\eqref{eq:HFE} is not a simple task. Due to the two timescales, the fast oscillation and the slower variation parameterised by $\lambda$, standard techniques for the study of Floquet systems are not directly applicable. Some approaches could be adapted, e.g., by obtaining the Floquet modes and quasienergies around fixed $\lambda$ and taking advantage of the slow variation of $\lambda$ with respect to the fast oscillation. However, our aim in this work is to demonstrate the potential of the proposed CAFFEINE method, therefore, we will focus on limits of modest $N$ and low $N_k$ where direct numerical simulation of the dynamics is possible. We also note that the final states from the dual annealing optimisation performed for CAFFEINE presented in this section have in the most part ended due to running out of iterations. Increasing the global search iterations and the maximum function evaluations could result in improvement of the results. This is especially the case when considering $N_k>1$, for which the dynamical simulations are time intensive.

First we consider the case of $N_k=1$ for various system sizes in Fig.~\ref{fig:Fig5}a and compare this to the case of unassisted annealing. We observe for even the simplest case of $N_k=1$ and $N_\tau=1$, i.e., $\beta_1 = \mathrm{const.}$, we observe a substantial reduction in the final state energy across the full range of $N$ studied. This can then be further improved by allowing for a more varied time-dependent drive of the highly oscillating terms by increasing $N_\tau$. We would not expect CAFFEINE with $N_k=1$ to function as well for large systems as it does for smaller systems, and we do observe that with a general increase in the distance of the final energy from the target energy. We note that for CAFFEINE this does not scale directly proportional to $N$ as is the case for the unassisted protocol, where the dynamics is frozen.

In addition we consider two cases of fixed size, $N=4$ and $N=6$, with increasing $N_k$ in Fig.~\ref{fig:Fig5}b and c respectively. We observe, as expected, an improved final state energy for increasing $N_k$ and $N_\tau$. The slight increase in final energy for the case of $N_k=4$ and $N_\tau=3$ can not be read into significantly, as the optimisation protocol ended due to reaching the maximum number of global iterations allowed.

\section{Conclusions}

We have introduced CAFFEINE, a new protocol for quantum annealing and state preparation problems. This protocol utilises findings from the field of counterdiabatic driving, mainly that it can be emulated via a Floquet engineered Hamiltonian. By parameterising this Floquet Hamiltonian then utilising standard optimisiation protocols we have shown that it is possible to construct small entangeld states with high fidelity and to improve quantum annealling protocols. We have also shown that it is possible to utilise this approach to learn the adiabatic gauge potential (AGP), potentially opening new avenues for obtaining the AGP from quantum hardware. This is particularly interesting due to the connections between the AGP and geometric approaches to quantum dynamics, including the dynamical crossing of a phase transition and chaotic systems.

For many control protocols the barrier to practical implementation is the requirement of additional control fields that may not be accessible or physically possible for a given experiment. CAFFEINE has been specifically designed to avoid this barrier by only requiring control of terms that are already controlled in the Hamiltonian implemented for the unassisted case. However, CAFFEINE does assume a rather extreme degree of control is possible, that is the fast oscillation of all terms. The short driving timescales enabled by CAFFEINE do mean that the system does not need to be driven indefinitely and the protocol could be completed in principle in a time well below that where heating from periodic driving becomes an issue \cite{eckardt2017colloquium,Bukov2016heating,Abanin2015exponentially,Mori2016rigorous}.

Beyond its capabilities for state preparation, annealing, and learning the AGP, the introduction of CAFFEINE opens a number of other questions, including methods for simulating such protocols. While it is clear from the presented results that scaling up both the number of oscillating frequencies $N_k$ and the number of discrete timesteps for the coefficients $N_\tau$ in general will improve the final result. With the somewhat impractical expectation that as both tend to $\infty$ the exact counterdiabatic dynamics should be obtained with perfect adiabatic evolution in arbitrary times. However, as we increase $N_k$ we are not just introducing new orthogonal terms but also corrections to the prior terms resulting in a more complex cost function landscape. This issue is well known from the commutator form of the AGP given by Eq.~\eqref{eq:AGP}, and is normally corrected by othogonallising each new term. It is not clear how to overcome this issue with non-orthogonal terms as we scale up $N_k$ in CAFFEINE but it could be crucial for future implementation. It could also be fuitful to investigate links between CAFFEINE and the separately considered topics of CD for periodically driven systems \cite{Schindler2024counterdiabtic} and the recent derivation of Floquet theory from the AGP \cite{schindler2024geometric}.

\acknowledgements{Work at the University of Strathclyde was supported by the Engineering and Physical Sciences Research Council through Grant No. EP/Y005058/2.}


\begin{thebibliography}{63}%
	\makeatletter
	\providecommand \@ifxundefined [1]{%
	 \@ifx{#1\undefined}
	}%
	\providecommand \@ifnum [1]{%
	 \ifnum #1\expandafter \@firstoftwo
	 \else \expandafter \@secondoftwo
	 \fi
	}%
	\providecommand \@ifx [1]{%
	 \ifx #1\expandafter \@firstoftwo
	 \else \expandafter \@secondoftwo
	 \fi
	}%
	\providecommand \natexlab [1]{#1}%
	\providecommand \enquote  [1]{``#1''}%
	\providecommand \bibnamefont  [1]{#1}%
	\providecommand \bibfnamefont [1]{#1}%
	\providecommand \citenamefont [1]{#1}%
	\providecommand \href@noop [0]{\@secondoftwo}%
	\providecommand \href [0]{\begingroup \@sanitize@url \@href}%
	\providecommand \@href[1]{\@@startlink{#1}\@@href}%
	\providecommand \@@href[1]{\endgroup#1\@@endlink}%
	\providecommand \@sanitize@url [0]{\catcode `\\12\catcode `\$12\catcode `\&12\catcode `\#12\catcode `\^12\catcode `\_12\catcode `\%12\relax}%
	\providecommand \@@startlink[1]{}%
	\providecommand \@@endlink[0]{}%
	\providecommand \url  [0]{\begingroup\@sanitize@url \@url }%
	\providecommand \@url [1]{\endgroup\@href {#1}{\urlprefix }}%
	\providecommand \urlprefix  [0]{URL }%
	\providecommand \Eprint [0]{\href }%
	\providecommand \doibase [0]{https://doi.org/}%
	\providecommand \selectlanguage [0]{\@gobble}%
	\providecommand \bibinfo  [0]{\@secondoftwo}%
	\providecommand \bibfield  [0]{\@secondoftwo}%
	\providecommand \translation [1]{[#1]}%
	\providecommand \BibitemOpen [0]{}%
	\providecommand \bibitemStop [0]{}%
	\providecommand \bibitemNoStop [0]{.\EOS\space}%
	\providecommand \EOS [0]{\spacefactor3000\relax}%
	\providecommand \BibitemShut  [1]{\csname bibitem#1\endcsname}%
	\let\auto@bib@innerbib\@empty
	\bibitem [{\citenamefont {Aharonov}\ \emph {et~al.}(2008)\citenamefont {Aharonov}, \citenamefont {Van~Dam}, \citenamefont {Kempe}, \citenamefont {Landau}, \citenamefont {Lloyd},\ and\ \citenamefont {Regev}}]{aharonov2008adiabatic}%
	  \BibitemOpen
	  \bibfield  {author} {\bibinfo {author} {\bibfnamefont {D.}~\bibnamefont {Aharonov}}, \bibinfo {author} {\bibfnamefont {W.}~\bibnamefont {Van~Dam}}, \bibinfo {author} {\bibfnamefont {J.}~\bibnamefont {Kempe}}, \bibinfo {author} {\bibfnamefont {Z.}~\bibnamefont {Landau}}, \bibinfo {author} {\bibfnamefont {S.}~\bibnamefont {Lloyd}},\ and\ \bibinfo {author} {\bibfnamefont {O.}~\bibnamefont {Regev}},\ }\href {https://doi.org/10.1137/080734479} {\bibfield  {journal} {\bibinfo  {journal} {SIAM review}\ }\textbf {\bibinfo {volume} {50}},\ \bibinfo {pages} {755} (\bibinfo {year} {2008})}\BibitemShut {NoStop}%
	\bibitem [{\citenamefont {Albash}\ and\ \citenamefont {Lidar}(2018)}]{Ablash2018adiabatic}%
	  \BibitemOpen
	  \bibfield  {author} {\bibinfo {author} {\bibfnamefont {T.}~\bibnamefont {Albash}}\ and\ \bibinfo {author} {\bibfnamefont {D.~A.}\ \bibnamefont {Lidar}},\ }\href {https://doi.org/10.1103/RevModPhys.90.015002} {\bibfield  {journal} {\bibinfo  {journal} {Rev. Mod. Phys.}\ }\textbf {\bibinfo {volume} {90}},\ \bibinfo {pages} {015002} (\bibinfo {year} {2018})}\BibitemShut {NoStop}%
	\bibitem [{\citenamefont {Morita}\ and\ \citenamefont {Nishimori}(2008)}]{morita2008mathematical}%
	  \BibitemOpen
	  \bibfield  {author} {\bibinfo {author} {\bibfnamefont {S.}~\bibnamefont {Morita}}\ and\ \bibinfo {author} {\bibfnamefont {H.}~\bibnamefont {Nishimori}},\ }\bibfield  {journal} {\bibinfo  {journal} {J. Math. Phys.}\ }\textbf {\bibinfo {volume} {49}},\ \href {https://doi.org/10.1063/1.2995837} {10.1063/1.2995837} (\bibinfo {year} {2008})\BibitemShut {NoStop}%
	\bibitem [{\citenamefont {Yarkoni}\ \emph {et~al.}(2022)\citenamefont {Yarkoni}, \citenamefont {Raponi}, \citenamefont {B{\"a}ck},\ and\ \citenamefont {Schmitt}}]{yarkoni2022quantum}%
	  \BibitemOpen
	  \bibfield  {author} {\bibinfo {author} {\bibfnamefont {S.}~\bibnamefont {Yarkoni}}, \bibinfo {author} {\bibfnamefont {E.}~\bibnamefont {Raponi}}, \bibinfo {author} {\bibfnamefont {T.}~\bibnamefont {B{\"a}ck}},\ and\ \bibinfo {author} {\bibfnamefont {S.}~\bibnamefont {Schmitt}},\ }\href {https://doi.org/10.1088/1361-6633/ac8c54} {\bibfield  {journal} {\bibinfo  {journal} {Rep. Prog. Phys.}\ }\textbf {\bibinfo {volume} {85}},\ \bibinfo {pages} {104001} (\bibinfo {year} {2022})}\BibitemShut {NoStop}%
	\bibitem [{\citenamefont {Rajak}\ \emph {et~al.}(2023)\citenamefont {Rajak}, \citenamefont {Suzuki}, \citenamefont {Dutta},\ and\ \citenamefont {Chakrabarti}}]{rajak2023quantum}%
	  \BibitemOpen
	  \bibfield  {author} {\bibinfo {author} {\bibfnamefont {A.}~\bibnamefont {Rajak}}, \bibinfo {author} {\bibfnamefont {S.}~\bibnamefont {Suzuki}}, \bibinfo {author} {\bibfnamefont {A.}~\bibnamefont {Dutta}},\ and\ \bibinfo {author} {\bibfnamefont {B.~K.}\ \bibnamefont {Chakrabarti}},\ }\href {https://doi.org/10.1098/rsta.2021.0417} {\bibfield  {journal} {\bibinfo  {journal} {Philos. T. R. Soc. A}\ }\textbf {\bibinfo {volume} {381}},\ \bibinfo {pages} {20210417} (\bibinfo {year} {2023})}\BibitemShut {NoStop}%
	\bibitem [{\citenamefont {Born}\ and\ \citenamefont {Fock}(1928)}]{born1928beweis}%
	  \BibitemOpen
	  \bibfield  {author} {\bibinfo {author} {\bibfnamefont {M.}~\bibnamefont {Born}}\ and\ \bibinfo {author} {\bibfnamefont {V.}~\bibnamefont {Fock}},\ }\href {https://doi.org/10.1007/BF01343193} {\bibfield  {journal} {\bibinfo  {journal} {Zeitschrift f{\"u}r Physik}\ }\textbf {\bibinfo {volume} {51}},\ \bibinfo {pages} {165} (\bibinfo {year} {1928})}\BibitemShut {NoStop}%
	\bibitem [{\citenamefont {Kato}(1950)}]{kato1950adiabatic}%
	  \BibitemOpen
	  \bibfield  {author} {\bibinfo {author} {\bibfnamefont {T.}~\bibnamefont {Kato}},\ }\href {https://doi.org/10.1143/JPSJ.5.435} {\bibfield  {journal} {\bibinfo  {journal} {J. Phys. Soc. Jpn.}\ }\textbf {\bibinfo {volume} {5}},\ \bibinfo {pages} {435} (\bibinfo {year} {1950})}\BibitemShut {NoStop}%
	\bibitem [{\citenamefont {Jansen}\ \emph {et~al.}(2007)\citenamefont {Jansen}, \citenamefont {Ruskai},\ and\ \citenamefont {Seiler}}]{jansen2007bounds}%
	  \BibitemOpen
	  \bibfield  {author} {\bibinfo {author} {\bibfnamefont {S.}~\bibnamefont {Jansen}}, \bibinfo {author} {\bibfnamefont {M.-B.}\ \bibnamefont {Ruskai}},\ and\ \bibinfo {author} {\bibfnamefont {R.}~\bibnamefont {Seiler}},\ }\bibfield  {journal} {\bibinfo  {journal} {J. Math. Phys.}\ }\textbf {\bibinfo {volume} {48}},\ \href {https://doi.org/10.1063/1.2798382} {10.1063/1.2798382} (\bibinfo {year} {2007})\BibitemShut {NoStop}%
	\bibitem [{\citenamefont {Gu\'ery-Odelin}\ \emph {et~al.}(2019)\citenamefont {Gu\'ery-Odelin}, \citenamefont {Ruschhaupt}, \citenamefont {Kiely}, \citenamefont {Torrontegui}, \citenamefont {Mart\'{\i}nez-Garaot},\ and\ \citenamefont {Muga}}]{gueryodelin2019shortcuts}%
	  \BibitemOpen
	  \bibfield  {author} {\bibinfo {author} {\bibfnamefont {D.}~\bibnamefont {Gu\'ery-Odelin}}, \bibinfo {author} {\bibfnamefont {A.}~\bibnamefont {Ruschhaupt}}, \bibinfo {author} {\bibfnamefont {A.}~\bibnamefont {Kiely}}, \bibinfo {author} {\bibfnamefont {E.}~\bibnamefont {Torrontegui}}, \bibinfo {author} {\bibfnamefont {S.}~\bibnamefont {Mart\'{\i}nez-Garaot}},\ and\ \bibinfo {author} {\bibfnamefont {J.~G.}\ \bibnamefont {Muga}},\ }\href {https://doi.org/10.1103/RevModPhys.91.045001} {\bibfield  {journal} {\bibinfo  {journal} {Rev. Mod. Phys.}\ }\textbf {\bibinfo {volume} {91}},\ \bibinfo {pages} {045001} (\bibinfo {year} {2019})}\BibitemShut {NoStop}%
	\bibitem [{\citenamefont {Torrontegui}\ \emph {et~al.}(2013)\citenamefont {Torrontegui}, \citenamefont {Ib{\'a}{\~n}ez}, \citenamefont {Mart{\'\i}nez-Garaot}, \citenamefont {Modugno}, \citenamefont {del Campo}, \citenamefont {Gu{\'e}ry-Odelin}, \citenamefont {Ruschhaupt}, \citenamefont {Chen},\ and\ \citenamefont {Muga}}]{torrontegui2013shortcuts}%
	  \BibitemOpen
	  \bibfield  {author} {\bibinfo {author} {\bibfnamefont {E.}~\bibnamefont {Torrontegui}}, \bibinfo {author} {\bibfnamefont {S.}~\bibnamefont {Ib{\'a}{\~n}ez}}, \bibinfo {author} {\bibfnamefont {S.}~\bibnamefont {Mart{\'\i}nez-Garaot}}, \bibinfo {author} {\bibfnamefont {M.}~\bibnamefont {Modugno}}, \bibinfo {author} {\bibfnamefont {A.}~\bibnamefont {del Campo}}, \bibinfo {author} {\bibfnamefont {D.}~\bibnamefont {Gu{\'e}ry-Odelin}}, \bibinfo {author} {\bibfnamefont {A.}~\bibnamefont {Ruschhaupt}}, \bibinfo {author} {\bibfnamefont {X.}~\bibnamefont {Chen}},\ and\ \bibinfo {author} {\bibfnamefont {J.~G.}\ \bibnamefont {Muga}},\ }in\ \href {https://doi.org/10.1016/B978-0-12-408090-4.00002-5} {\emph {\bibinfo {booktitle} {Adv. Atom Mol. Opt. Phys.}}},\ Vol.~\bibinfo {volume} {62}\ (\bibinfo  {publisher} {Elsevier},\ \bibinfo {year} {2013})\ pp.\ \bibinfo {pages} {117--169}\BibitemShut {NoStop}%
	\bibitem [{\citenamefont {Glaser}\ \emph {et~al.}(2015)\citenamefont {Glaser}, \citenamefont {Boscain}, \citenamefont {Calarco}, \citenamefont {Koch}, \citenamefont {K{\"o}ckenberger}, \citenamefont {Kosloff}, \citenamefont {Kuprov}, \citenamefont {Luy}, \citenamefont {Schirmer}, \citenamefont {Schulte-Herbr{\"u}ggen} \emph {et~al.}}]{glaser2015training}%
	  \BibitemOpen
	  \bibfield  {author} {\bibinfo {author} {\bibfnamefont {S.~J.}\ \bibnamefont {Glaser}}, \bibinfo {author} {\bibfnamefont {U.}~\bibnamefont {Boscain}}, \bibinfo {author} {\bibfnamefont {T.}~\bibnamefont {Calarco}}, \bibinfo {author} {\bibfnamefont {C.~P.}\ \bibnamefont {Koch}}, \bibinfo {author} {\bibfnamefont {W.}~\bibnamefont {K{\"o}ckenberger}}, \bibinfo {author} {\bibfnamefont {R.}~\bibnamefont {Kosloff}}, \bibinfo {author} {\bibfnamefont {I.}~\bibnamefont {Kuprov}}, \bibinfo {author} {\bibfnamefont {B.}~\bibnamefont {Luy}}, \bibinfo {author} {\bibfnamefont {S.}~\bibnamefont {Schirmer}}, \bibinfo {author} {\bibfnamefont {T.}~\bibnamefont {Schulte-Herbr{\"u}ggen}}, \emph {et~al.},\ }\href {https://doi.org/10.1140/epjd/e2015-60464-1} {\bibfield  {journal} {\bibinfo  {journal} {Eur. Phys. J. D}\ }\textbf {\bibinfo {volume} {69}},\ \bibinfo {pages} {1} (\bibinfo {year} {2015})}\BibitemShut {NoStop}%
	\bibitem [{\citenamefont {Ansel}\ \emph {et~al.}(2024)\citenamefont {Ansel}, \citenamefont {Dionis}, \citenamefont {Arrouas}, \citenamefont {Peaudecerf}, \citenamefont {Gu{\'e}rin}, \citenamefont {Gu{\'e}ry-Odelin},\ and\ \citenamefont {Sugny}}]{ansel2024introduction}%
	  \BibitemOpen
	  \bibfield  {author} {\bibinfo {author} {\bibfnamefont {Q.}~\bibnamefont {Ansel}}, \bibinfo {author} {\bibfnamefont {E.}~\bibnamefont {Dionis}}, \bibinfo {author} {\bibfnamefont {F.}~\bibnamefont {Arrouas}}, \bibinfo {author} {\bibfnamefont {B.}~\bibnamefont {Peaudecerf}}, \bibinfo {author} {\bibfnamefont {S.}~\bibnamefont {Gu{\'e}rin}}, \bibinfo {author} {\bibfnamefont {D.}~\bibnamefont {Gu{\'e}ry-Odelin}},\ and\ \bibinfo {author} {\bibfnamefont {D.}~\bibnamefont {Sugny}},\ }\href {https://doi.org/10.1088/1361-6455/ad46a5} {\bibfield  {journal} {\bibinfo  {journal} {J. Phys. B: At. Mol. Opt.}\ }\textbf {\bibinfo {volume} {57}},\ \bibinfo {pages} {133001} (\bibinfo {year} {2024})}\BibitemShut {NoStop}%
	\bibitem [{\citenamefont {Demirplak}\ and\ \citenamefont {Rice}(2003)}]{demirplak2003adiabatic}%
	  \BibitemOpen
	  \bibfield  {author} {\bibinfo {author} {\bibfnamefont {M.}~\bibnamefont {Demirplak}}\ and\ \bibinfo {author} {\bibfnamefont {S.~A.}\ \bibnamefont {Rice}},\ }\href {https://doi.org/10.1021/jp030708a} {\bibfield  {journal} {\bibinfo  {journal} {The Journal of Physical Chemistry A}\ }\textbf {\bibinfo {volume} {107}},\ \bibinfo {pages} {9937} (\bibinfo {year} {2003})}\BibitemShut {NoStop}%
	\bibitem [{\citenamefont {Demirplak}\ and\ \citenamefont {Rice}(2005)}]{demirplak2005assisted}%
	  \BibitemOpen
	  \bibfield  {author} {\bibinfo {author} {\bibfnamefont {M.}~\bibnamefont {Demirplak}}\ and\ \bibinfo {author} {\bibfnamefont {S.~A.}\ \bibnamefont {Rice}},\ }\href {https://doi.org/10.1021/jp040647w} {\bibfield  {journal} {\bibinfo  {journal} {The Journal of Physical Chemistry B}\ }\textbf {\bibinfo {volume} {109}},\ \bibinfo {pages} {6838} (\bibinfo {year} {2005})}\BibitemShut {NoStop}%
	\bibitem [{\citenamefont {Berry}(2009)}]{berry2009transitionless}%
	  \BibitemOpen
	  \bibfield  {author} {\bibinfo {author} {\bibfnamefont {M.~V.}\ \bibnamefont {Berry}},\ }\href {https://doi.org/10.1088/1751-8113/42/36/365303} {\bibfield  {journal} {\bibinfo  {journal} {Journal of Physics A: Mathematical and Theoretical}\ }\textbf {\bibinfo {volume} {42}},\ \bibinfo {pages} {365303} (\bibinfo {year} {2009})}\BibitemShut {NoStop}%
	\bibitem [{\citenamefont {del Campo}(2013)}]{campo2013shortcuts}%
	  \BibitemOpen
	  \bibfield  {author} {\bibinfo {author} {\bibfnamefont {A.}~\bibnamefont {del Campo}},\ }\href {https://doi.org/10.1103/PhysRevLett.111.100502} {\bibfield  {journal} {\bibinfo  {journal} {Phys. Rev. Lett.}\ }\textbf {\bibinfo {volume} {111}},\ \bibinfo {pages} {100502} (\bibinfo {year} {2013})}\BibitemShut {NoStop}%
	\bibitem [{\citenamefont {del Campo}\ \emph {et~al.}(2012)\citenamefont {del Campo}, \citenamefont {Rams},\ and\ \citenamefont {Zurek}}]{campo2012assisted}%
	  \BibitemOpen
	  \bibfield  {author} {\bibinfo {author} {\bibfnamefont {A.}~\bibnamefont {del Campo}}, \bibinfo {author} {\bibfnamefont {M.~M.}\ \bibnamefont {Rams}},\ and\ \bibinfo {author} {\bibfnamefont {W.~H.}\ \bibnamefont {Zurek}},\ }\href {https://doi.org/10.1103/PhysRevLett.109.115703} {\bibfield  {journal} {\bibinfo  {journal} {Phys. Rev. Lett.}\ }\textbf {\bibinfo {volume} {109}},\ \bibinfo {pages} {115703} (\bibinfo {year} {2012})}\BibitemShut {NoStop}%
	\bibitem [{\citenamefont {Damski}(2014)}]{damski2014counterdiabatic}%
	  \BibitemOpen
	  \bibfield  {author} {\bibinfo {author} {\bibfnamefont {B.}~\bibnamefont {Damski}},\ }\href {https://doi.org/10.1088/1742-5468/2014/12/P12019} {\bibfield  {journal} {\bibinfo  {journal} {J. Stat. Mech.: Theory E.}\ }\textbf {\bibinfo {volume} {2014}},\ \bibinfo {pages} {P12019} (\bibinfo {year} {2014})}\BibitemShut {NoStop}%
	\bibitem [{\citenamefont {Duncan}(2024)}]{duncan2024exact}%
	  \BibitemOpen
	  \bibfield  {author} {\bibinfo {author} {\bibfnamefont {C.~W.}\ \bibnamefont {Duncan}},\ }\href {https://doi.org/10.1103/PhysRevB.109.245421} {\bibfield  {journal} {\bibinfo  {journal} {Phys. Rev. B}\ }\textbf {\bibinfo {volume} {109}},\ \bibinfo {pages} {245421} (\bibinfo {year} {2024})}\BibitemShut {NoStop}%
	\bibitem [{\citenamefont {Lawrence}\ \emph {et~al.}(2025)\citenamefont {Lawrence}, \citenamefont {Schmid}, \citenamefont {{\v{C}}epait{\.e}}, \citenamefont {Kirton},\ and\ \citenamefont {Duncan}}]{lawrence2024numerical}%
	  \BibitemOpen
	  \bibfield  {author} {\bibinfo {author} {\bibfnamefont {E.}~\bibnamefont {Lawrence}}, \bibinfo {author} {\bibfnamefont {S.~F.}\ \bibnamefont {Schmid}}, \bibinfo {author} {\bibfnamefont {I.}~\bibnamefont {{\v{C}}epait{\.e}}}, \bibinfo {author} {\bibfnamefont {P.}~\bibnamefont {Kirton}},\ and\ \bibinfo {author} {\bibfnamefont {C.~W.}\ \bibnamefont {Duncan}},\ }\href {https://doi.org/10.21468/SciPostPhys.18.1.014} {\bibfield  {journal} {\bibinfo  {journal} {SciPost Physics}\ }\textbf {\bibinfo {volume} {18}},\ \bibinfo {pages} {014} (\bibinfo {year} {2025})}\BibitemShut {NoStop}%
	\bibitem [{\citenamefont {Takahashi}\ and\ \citenamefont {del Campo}(2024)}]{Takahashi2024shortcuts}%
	  \BibitemOpen
	  \bibfield  {author} {\bibinfo {author} {\bibfnamefont {K.}~\bibnamefont {Takahashi}}\ and\ \bibinfo {author} {\bibfnamefont {A.}~\bibnamefont {del Campo}},\ }\href {https://doi.org/10.1103/PhysRevX.14.011032} {\bibfield  {journal} {\bibinfo  {journal} {Phys. Rev. X}\ }\textbf {\bibinfo {volume} {14}},\ \bibinfo {pages} {011032} (\bibinfo {year} {2024})}\BibitemShut {NoStop}%
	\bibitem [{\citenamefont {Sels}\ and\ \citenamefont {Polkovnikov}(2017)}]{sels2017minimizing}%
	  \BibitemOpen
	  \bibfield  {author} {\bibinfo {author} {\bibfnamefont {D.}~\bibnamefont {Sels}}\ and\ \bibinfo {author} {\bibfnamefont {A.}~\bibnamefont {Polkovnikov}},\ }\href {https://doi.org/10.1073/pnas.1619826114} {\bibfield  {journal} {\bibinfo  {journal} {Proceedings of the National Academy of Sciences}\ }\textbf {\bibinfo {volume} {114}},\ \bibinfo {pages} {E3909} (\bibinfo {year} {2017})}\BibitemShut {NoStop}%
	\bibitem [{\citenamefont {Kolodrubetz}\ \emph {et~al.}(2017)\citenamefont {Kolodrubetz}, \citenamefont {Sels}, \citenamefont {Mehta},\ and\ \citenamefont {Polkovnikov}}]{kolodrubetz2017geometry}%
	  \BibitemOpen
	  \bibfield  {author} {\bibinfo {author} {\bibfnamefont {M.}~\bibnamefont {Kolodrubetz}}, \bibinfo {author} {\bibfnamefont {D.}~\bibnamefont {Sels}}, \bibinfo {author} {\bibfnamefont {P.}~\bibnamefont {Mehta}},\ and\ \bibinfo {author} {\bibfnamefont {A.}~\bibnamefont {Polkovnikov}},\ }\href {https://doi.org/10.1016/j.physrep.2017.07.001} {\bibfield  {journal} {\bibinfo  {journal} {Phys. Rep.}\ }\textbf {\bibinfo {volume} {697}},\ \bibinfo {pages} {1} (\bibinfo {year} {2017})}\BibitemShut {NoStop}%
	\bibitem [{\citenamefont {van Vreumingen}(2024)}]{Vreumingen2024gate}%
	  \BibitemOpen
	  \bibfield  {author} {\bibinfo {author} {\bibfnamefont {D.}~\bibnamefont {van Vreumingen}},\ }\href {https://doi.org/10.1103/PhysRevA.110.052419} {\bibfield  {journal} {\bibinfo  {journal} {Phys. Rev. A}\ }\textbf {\bibinfo {volume} {110}},\ \bibinfo {pages} {052419} (\bibinfo {year} {2024})}\BibitemShut {NoStop}%
	\bibitem [{\citenamefont {Li}\ \emph {et~al.}(2024)\citenamefont {Li}, \citenamefont {Shen}, \citenamefont {Shaydulin},\ and\ \citenamefont {Pistoia}}]{li2024quantum}%
	  \BibitemOpen
	  \bibfield  {author} {\bibinfo {author} {\bibfnamefont {C.}~\bibnamefont {Li}}, \bibinfo {author} {\bibfnamefont {J.}~\bibnamefont {Shen}}, \bibinfo {author} {\bibfnamefont {R.}~\bibnamefont {Shaydulin}},\ and\ \bibinfo {author} {\bibfnamefont {M.}~\bibnamefont {Pistoia}},\ }\bibfield  {journal} {\bibinfo  {journal} {arXiv preprint arXiv:2403.01854}\ }\href {https://doi.org/10.48550/arXiv.2403.01854} {10.48550/arXiv.2403.01854} (\bibinfo {year} {2024})\BibitemShut {NoStop}%
	\bibitem [{\citenamefont {Hegade}\ \emph {et~al.}(2022)\citenamefont {Hegade}, \citenamefont {Chen},\ and\ \citenamefont {Solano}}]{Hegade2022digitized}%
	  \BibitemOpen
	  \bibfield  {author} {\bibinfo {author} {\bibfnamefont {N.~N.}\ \bibnamefont {Hegade}}, \bibinfo {author} {\bibfnamefont {X.}~\bibnamefont {Chen}},\ and\ \bibinfo {author} {\bibfnamefont {E.}~\bibnamefont {Solano}},\ }\href {https://doi.org/10.1103/PhysRevResearch.4.L042030} {\bibfield  {journal} {\bibinfo  {journal} {Phys. Rev. Res.}\ }\textbf {\bibinfo {volume} {4}},\ \bibinfo {pages} {L042030} (\bibinfo {year} {2022})}\BibitemShut {NoStop}%
	\bibitem [{\citenamefont {Wurtz}\ and\ \citenamefont {Love}(2022)}]{wurtz2022counterdiabaticity}%
	  \BibitemOpen
	  \bibfield  {author} {\bibinfo {author} {\bibfnamefont {J.}~\bibnamefont {Wurtz}}\ and\ \bibinfo {author} {\bibfnamefont {P.~J.}\ \bibnamefont {Love}},\ }\href {https://doi.org/10.22331/q-2022-01-27-635} {\bibfield  {journal} {\bibinfo  {journal} {Quantum}\ }\textbf {\bibinfo {volume} {6}},\ \bibinfo {pages} {635} (\bibinfo {year} {2022})}\BibitemShut {NoStop}%
	\bibitem [{\citenamefont {Chandarana}\ \emph {et~al.}(2022)\citenamefont {Chandarana}, \citenamefont {Hegade}, \citenamefont {Paul}, \citenamefont {Albarr\'an-Arriagada}, \citenamefont {Solano}, \citenamefont {del Campo},\ and\ \citenamefont {Chen}}]{Chandarana2022digitized}%
	  \BibitemOpen
	  \bibfield  {author} {\bibinfo {author} {\bibfnamefont {P.}~\bibnamefont {Chandarana}}, \bibinfo {author} {\bibfnamefont {N.~N.}\ \bibnamefont {Hegade}}, \bibinfo {author} {\bibfnamefont {K.}~\bibnamefont {Paul}}, \bibinfo {author} {\bibfnamefont {F.}~\bibnamefont {Albarr\'an-Arriagada}}, \bibinfo {author} {\bibfnamefont {E.}~\bibnamefont {Solano}}, \bibinfo {author} {\bibfnamefont {A.}~\bibnamefont {del Campo}},\ and\ \bibinfo {author} {\bibfnamefont {X.}~\bibnamefont {Chen}},\ }\href {https://doi.org/10.1103/PhysRevResearch.4.013141} {\bibfield  {journal} {\bibinfo  {journal} {Phys. Rev. Res.}\ }\textbf {\bibinfo {volume} {4}},\ \bibinfo {pages} {013141} (\bibinfo {year} {2022})}\BibitemShut {NoStop}%
	\bibitem [{\citenamefont {Blekos}\ \emph {et~al.}(2024)\citenamefont {Blekos}, \citenamefont {Brand}, \citenamefont {Ceschini}, \citenamefont {Chou}, \citenamefont {Li}, \citenamefont {Pandya},\ and\ \citenamefont {Summer}}]{blekos2024review}%
	  \BibitemOpen
	  \bibfield  {author} {\bibinfo {author} {\bibfnamefont {K.}~\bibnamefont {Blekos}}, \bibinfo {author} {\bibfnamefont {D.}~\bibnamefont {Brand}}, \bibinfo {author} {\bibfnamefont {A.}~\bibnamefont {Ceschini}}, \bibinfo {author} {\bibfnamefont {C.-H.}\ \bibnamefont {Chou}}, \bibinfo {author} {\bibfnamefont {R.-H.}\ \bibnamefont {Li}}, \bibinfo {author} {\bibfnamefont {K.}~\bibnamefont {Pandya}},\ and\ \bibinfo {author} {\bibfnamefont {A.}~\bibnamefont {Summer}},\ }\href {https://doi.org/10.1016/j.physrep.2024.03.002} {\bibfield  {journal} {\bibinfo  {journal} {Phys. Rep.}\ }\textbf {\bibinfo {volume} {1068}},\ \bibinfo {pages} {1} (\bibinfo {year} {2024})}\BibitemShut {NoStop}%
	\bibitem [{\citenamefont {Malla}\ \emph {et~al.}(2024)\citenamefont {Malla}, \citenamefont {Sukeno}, \citenamefont {Yu}, \citenamefont {Wei}, \citenamefont {Weichselbaum},\ and\ \citenamefont {Konik}}]{malla2024feedback}%
	  \BibitemOpen
	  \bibfield  {author} {\bibinfo {author} {\bibfnamefont {R.~K.}\ \bibnamefont {Malla}}, \bibinfo {author} {\bibfnamefont {H.}~\bibnamefont {Sukeno}}, \bibinfo {author} {\bibfnamefont {H.}~\bibnamefont {Yu}}, \bibinfo {author} {\bibfnamefont {T.-C.}\ \bibnamefont {Wei}}, \bibinfo {author} {\bibfnamefont {A.}~\bibnamefont {Weichselbaum}},\ and\ \bibinfo {author} {\bibfnamefont {R.~M.}\ \bibnamefont {Konik}},\ }\href {https://doi.org/10.1103/PhysRevResearch.6.043068} {\bibfield  {journal} {\bibinfo  {journal} {Phys. Rev. Res.}\ }\textbf {\bibinfo {volume} {6}},\ \bibinfo {pages} {043068} (\bibinfo {year} {2024})}\BibitemShut {NoStop}%
	\bibitem [{\citenamefont {Saberi}\ \emph {et~al.}(2014)\citenamefont {Saberi}, \citenamefont {Opatrn\'y}, \citenamefont {M\o{}lmer},\ and\ \citenamefont {del Campo}}]{Saberi2014adiabatic}%
	  \BibitemOpen
	  \bibfield  {author} {\bibinfo {author} {\bibfnamefont {H.}~\bibnamefont {Saberi}}, \bibinfo {author} {\bibfnamefont {T.~c.~v.}\ \bibnamefont {Opatrn\'y}}, \bibinfo {author} {\bibfnamefont {K.}~\bibnamefont {M\o{}lmer}},\ and\ \bibinfo {author} {\bibfnamefont {A.}~\bibnamefont {del Campo}},\ }\href {https://doi.org/10.1103/PhysRevA.90.060301} {\bibfield  {journal} {\bibinfo  {journal} {Phys. Rev. A}\ }\textbf {\bibinfo {volume} {90}},\ \bibinfo {pages} {060301} (\bibinfo {year} {2014})}\BibitemShut {NoStop}%
	\bibitem [{\citenamefont {\ifmmode \check{C}\else \v{C}\fi{}epait\ifmmode~\dot{e}\else \.{e}\fi{}}\ \emph {et~al.}(2023)\citenamefont {\ifmmode \check{C}\else \v{C}\fi{}epait\ifmmode~\dot{e}\else \.{e}\fi{}}, \citenamefont {Polkovnikov}, \citenamefont {Daley},\ and\ \citenamefont {Duncan}}]{cepaite2023counterdiabatic}%
	  \BibitemOpen
	  \bibfield  {author} {\bibinfo {author} {\bibfnamefont {I.}~\bibnamefont {\ifmmode \check{C}\else \v{C}\fi{}epait\ifmmode~\dot{e}\else \.{e}\fi{}}}, \bibinfo {author} {\bibfnamefont {A.}~\bibnamefont {Polkovnikov}}, \bibinfo {author} {\bibfnamefont {A.~J.}\ \bibnamefont {Daley}},\ and\ \bibinfo {author} {\bibfnamefont {C.~W.}\ \bibnamefont {Duncan}},\ }\href {https://doi.org/10.1103/PRXQuantum.4.010312} {\bibfield  {journal} {\bibinfo  {journal} {PRX Quantum}\ }\textbf {\bibinfo {volume} {4}},\ \bibinfo {pages} {010312} (\bibinfo {year} {2023})}\BibitemShut {NoStop}%
	\bibitem [{\citenamefont {Morawetz}\ and\ \citenamefont {Polkovnikov}(2024)}]{morawetz2024efficient}%
	  \BibitemOpen
	  \bibfield  {author} {\bibinfo {author} {\bibfnamefont {S.}~\bibnamefont {Morawetz}}\ and\ \bibinfo {author} {\bibfnamefont {A.}~\bibnamefont {Polkovnikov}},\ }\href {https://doi.org/10.1103/PhysRevB.110.024304} {\bibfield  {journal} {\bibinfo  {journal} {Phys. Rev. B}\ }\textbf {\bibinfo {volume} {110}},\ \bibinfo {pages} {024304} (\bibinfo {year} {2024})}\BibitemShut {NoStop}%
	\bibitem [{\citenamefont {Petiziol}\ \emph {et~al.}(2018)\citenamefont {Petiziol}, \citenamefont {Dive}, \citenamefont {Mintert},\ and\ \citenamefont {Wimberger}}]{petiziol2018fast}%
	  \BibitemOpen
	  \bibfield  {author} {\bibinfo {author} {\bibfnamefont {F.}~\bibnamefont {Petiziol}}, \bibinfo {author} {\bibfnamefont {B.}~\bibnamefont {Dive}}, \bibinfo {author} {\bibfnamefont {F.}~\bibnamefont {Mintert}},\ and\ \bibinfo {author} {\bibfnamefont {S.}~\bibnamefont {Wimberger}},\ }\href {https://doi.org/10.1103/PhysRevA.98.043436} {\bibfield  {journal} {\bibinfo  {journal} {Phys. Rev. A}\ }\textbf {\bibinfo {volume} {98}},\ \bibinfo {pages} {043436} (\bibinfo {year} {2018})}\BibitemShut {NoStop}%
	\bibitem [{\citenamefont {Petiziol}\ \emph {et~al.}(2019)\citenamefont {Petiziol}, \citenamefont {Dive}, \citenamefont {Carretta}, \citenamefont {Mannella}, \citenamefont {Mintert},\ and\ \citenamefont {Wimberger}}]{petiziol2019accelerating}%
	  \BibitemOpen
	  \bibfield  {author} {\bibinfo {author} {\bibfnamefont {F.}~\bibnamefont {Petiziol}}, \bibinfo {author} {\bibfnamefont {B.}~\bibnamefont {Dive}}, \bibinfo {author} {\bibfnamefont {S.}~\bibnamefont {Carretta}}, \bibinfo {author} {\bibfnamefont {R.}~\bibnamefont {Mannella}}, \bibinfo {author} {\bibfnamefont {F.}~\bibnamefont {Mintert}},\ and\ \bibinfo {author} {\bibfnamefont {S.}~\bibnamefont {Wimberger}},\ }\href {https://doi.org/10.1103/PhysRevA.99.042315} {\bibfield  {journal} {\bibinfo  {journal} {Phys. Rev. A}\ }\textbf {\bibinfo {volume} {99}},\ \bibinfo {pages} {042315} (\bibinfo {year} {2019})}\BibitemShut {NoStop}%
	\bibitem [{\citenamefont {Villazon}\ \emph {et~al.}(2019)\citenamefont {Villazon}, \citenamefont {Polkovnikov},\ and\ \citenamefont {Chandran}}]{Villazon2019swift}%
	  \BibitemOpen
	  \bibfield  {author} {\bibinfo {author} {\bibfnamefont {T.}~\bibnamefont {Villazon}}, \bibinfo {author} {\bibfnamefont {A.}~\bibnamefont {Polkovnikov}},\ and\ \bibinfo {author} {\bibfnamefont {A.}~\bibnamefont {Chandran}},\ }\href {https://doi.org/10.1103/PhysRevA.100.012126} {\bibfield  {journal} {\bibinfo  {journal} {Phys. Rev. A}\ }\textbf {\bibinfo {volume} {100}},\ \bibinfo {pages} {012126} (\bibinfo {year} {2019})}\BibitemShut {NoStop}%
	\bibitem [{\citenamefont {Claeys}\ \emph {et~al.}(2019)\citenamefont {Claeys}, \citenamefont {Pandey}, \citenamefont {Sels},\ and\ \citenamefont {Polkovnikov}}]{claeys2019floquet}%
	  \BibitemOpen
	  \bibfield  {author} {\bibinfo {author} {\bibfnamefont {P.~W.}\ \bibnamefont {Claeys}}, \bibinfo {author} {\bibfnamefont {M.}~\bibnamefont {Pandey}}, \bibinfo {author} {\bibfnamefont {D.}~\bibnamefont {Sels}},\ and\ \bibinfo {author} {\bibfnamefont {A.}~\bibnamefont {Polkovnikov}},\ }\href {https://doi.org/10.1103/PhysRevLett.123.090602} {\bibfield  {journal} {\bibinfo  {journal} {Phys. Rev. Lett.}\ }\textbf {\bibinfo {volume} {123}},\ \bibinfo {pages} {090602} (\bibinfo {year} {2019})}\BibitemShut {NoStop}%
	\bibitem [{\citenamefont {Boyers}\ \emph {et~al.}(2019)\citenamefont {Boyers}, \citenamefont {Pandey}, \citenamefont {Campbell}, \citenamefont {Polkovnikov}, \citenamefont {Sels},\ and\ \citenamefont {Sushkov}}]{Boyers2019floquet}%
	  \BibitemOpen
	  \bibfield  {author} {\bibinfo {author} {\bibfnamefont {E.}~\bibnamefont {Boyers}}, \bibinfo {author} {\bibfnamefont {M.}~\bibnamefont {Pandey}}, \bibinfo {author} {\bibfnamefont {D.~K.}\ \bibnamefont {Campbell}}, \bibinfo {author} {\bibfnamefont {A.}~\bibnamefont {Polkovnikov}}, \bibinfo {author} {\bibfnamefont {D.}~\bibnamefont {Sels}},\ and\ \bibinfo {author} {\bibfnamefont {A.~O.}\ \bibnamefont {Sushkov}},\ }\href {https://doi.org/10.1103/PhysRevA.100.012341} {\bibfield  {journal} {\bibinfo  {journal} {Phys. Rev. A}\ }\textbf {\bibinfo {volume} {100}},\ \bibinfo {pages} {012341} (\bibinfo {year} {2019})}\BibitemShut {NoStop}%
	\bibitem [{\citenamefont {Zhou}\ \emph {et~al.}(2019)\citenamefont {Zhou}, \citenamefont {Chen}, \citenamefont {Nie}, \citenamefont {Bian}, \citenamefont {Ji}, \citenamefont {Li},\ and\ \citenamefont {Peng}}]{zhou2019floquet}%
	  \BibitemOpen
	  \bibfield  {author} {\bibinfo {author} {\bibfnamefont {H.}~\bibnamefont {Zhou}}, \bibinfo {author} {\bibfnamefont {X.}~\bibnamefont {Chen}}, \bibinfo {author} {\bibfnamefont {X.}~\bibnamefont {Nie}}, \bibinfo {author} {\bibfnamefont {J.}~\bibnamefont {Bian}}, \bibinfo {author} {\bibfnamefont {Y.}~\bibnamefont {Ji}}, \bibinfo {author} {\bibfnamefont {Z.}~\bibnamefont {Li}},\ and\ \bibinfo {author} {\bibfnamefont {X.}~\bibnamefont {Peng}},\ }\href {https://doi.org/10.1016/j.scib.2019.05.018} {\bibfield  {journal} {\bibinfo  {journal} {Science bulletin}\ }\textbf {\bibinfo {volume} {64}},\ \bibinfo {pages} {888} (\bibinfo {year} {2019})}\BibitemShut {NoStop}%
	\bibitem [{\citenamefont {Pandey}\ \emph {et~al.}(2020)\citenamefont {Pandey}, \citenamefont {Claeys}, \citenamefont {Campbell}, \citenamefont {Polkovnikov},\ and\ \citenamefont {Sels}}]{Pandey2020adiabatic}%
	  \BibitemOpen
	  \bibfield  {author} {\bibinfo {author} {\bibfnamefont {M.}~\bibnamefont {Pandey}}, \bibinfo {author} {\bibfnamefont {P.~W.}\ \bibnamefont {Claeys}}, \bibinfo {author} {\bibfnamefont {D.~K.}\ \bibnamefont {Campbell}}, \bibinfo {author} {\bibfnamefont {A.}~\bibnamefont {Polkovnikov}},\ and\ \bibinfo {author} {\bibfnamefont {D.}~\bibnamefont {Sels}},\ }\href {https://doi.org/10.1103/PhysRevX.10.041017} {\bibfield  {journal} {\bibinfo  {journal} {Phys. Rev. X}\ }\textbf {\bibinfo {volume} {10}},\ \bibinfo {pages} {041017} (\bibinfo {year} {2020})}\BibitemShut {NoStop}%
	\bibitem [{\citenamefont {Lim}\ \emph {et~al.}(2024)\citenamefont {Lim}, \citenamefont {Matirko}, \citenamefont {Polkovnikov},\ and\ \citenamefont {Flynn}}]{lim2024defining}%
	  \BibitemOpen
	  \bibfield  {author} {\bibinfo {author} {\bibfnamefont {C.}~\bibnamefont {Lim}}, \bibinfo {author} {\bibfnamefont {K.}~\bibnamefont {Matirko}}, \bibinfo {author} {\bibfnamefont {A.}~\bibnamefont {Polkovnikov}},\ and\ \bibinfo {author} {\bibfnamefont {M.~O.}\ \bibnamefont {Flynn}},\ }\bibfield  {journal} {\bibinfo  {journal} {arXiv preprint arXiv:2401.01927}\ }\href {https://doi.org/10.48550/arXiv.2401.01927} {10.48550/arXiv.2401.01927} (\bibinfo {year} {2024})\BibitemShut {NoStop}%
	\bibitem [{\citenamefont {Hatomura}\ and\ \citenamefont {Takahashi}(2021)}]{Hatomura2021controlling}%
	  \BibitemOpen
	  \bibfield  {author} {\bibinfo {author} {\bibfnamefont {T.}~\bibnamefont {Hatomura}}\ and\ \bibinfo {author} {\bibfnamefont {K.}~\bibnamefont {Takahashi}},\ }\href {https://doi.org/10.1103/PhysRevA.103.012220} {\bibfield  {journal} {\bibinfo  {journal} {Phys. Rev. A}\ }\textbf {\bibinfo {volume} {103}},\ \bibinfo {pages} {012220} (\bibinfo {year} {2021})}\BibitemShut {NoStop}%
	\bibitem [{\citenamefont {Balducci}\ \emph {et~al.}(2024)\citenamefont {Balducci}, \citenamefont {Grabarits},\ and\ \citenamefont {del Campo}}]{balducci2024fighting}%
	  \BibitemOpen
	  \bibfield  {author} {\bibinfo {author} {\bibfnamefont {F.}~\bibnamefont {Balducci}}, \bibinfo {author} {\bibfnamefont {A.}~\bibnamefont {Grabarits}},\ and\ \bibinfo {author} {\bibfnamefont {A.}~\bibnamefont {del Campo}},\ }\bibfield  {journal} {\bibinfo  {journal} {arXiv preprint arXiv:2410.02520}\ }\href {https://doi.org/10.48550/arXiv.2410.02520} {10.48550/arXiv.2410.02520} (\bibinfo {year} {2024})\BibitemShut {NoStop}%
	\bibitem [{\citenamefont {Caneva}\ \emph {et~al.}(2011)\citenamefont {Caneva}, \citenamefont {Calarco},\ and\ \citenamefont {Montangero}}]{Caneva2011chopped}%
	  \BibitemOpen
	  \bibfield  {author} {\bibinfo {author} {\bibfnamefont {T.}~\bibnamefont {Caneva}}, \bibinfo {author} {\bibfnamefont {T.}~\bibnamefont {Calarco}},\ and\ \bibinfo {author} {\bibfnamefont {S.}~\bibnamefont {Montangero}},\ }\href {https://doi.org/10.1103/PhysRevA.84.022326} {\bibfield  {journal} {\bibinfo  {journal} {Phys. Rev. A}\ }\textbf {\bibinfo {volume} {84}},\ \bibinfo {pages} {022326} (\bibinfo {year} {2011})}\BibitemShut {NoStop}%
	\bibitem [{\citenamefont {Rach}\ \emph {et~al.}(2015)\citenamefont {Rach}, \citenamefont {M\"uller}, \citenamefont {Calarco},\ and\ \citenamefont {Montangero}}]{Rach2015dressing}%
	  \BibitemOpen
	  \bibfield  {author} {\bibinfo {author} {\bibfnamefont {N.}~\bibnamefont {Rach}}, \bibinfo {author} {\bibfnamefont {M.~M.}\ \bibnamefont {M\"uller}}, \bibinfo {author} {\bibfnamefont {T.}~\bibnamefont {Calarco}},\ and\ \bibinfo {author} {\bibfnamefont {S.}~\bibnamefont {Montangero}},\ }\href {https://doi.org/10.1103/PhysRevA.92.062343} {\bibfield  {journal} {\bibinfo  {journal} {Phys. Rev. A}\ }\textbf {\bibinfo {volume} {92}},\ \bibinfo {pages} {062343} (\bibinfo {year} {2015})}\BibitemShut {NoStop}%
	\bibitem [{\citenamefont {M{\"u}ller}\ \emph {et~al.}(2022)\citenamefont {M{\"u}ller}, \citenamefont {Said}, \citenamefont {Jelezko}, \citenamefont {Calarco},\ and\ \citenamefont {Montangero}}]{muller2022one}%
	  \BibitemOpen
	  \bibfield  {author} {\bibinfo {author} {\bibfnamefont {M.~M.}\ \bibnamefont {M{\"u}ller}}, \bibinfo {author} {\bibfnamefont {R.~S.}\ \bibnamefont {Said}}, \bibinfo {author} {\bibfnamefont {F.}~\bibnamefont {Jelezko}}, \bibinfo {author} {\bibfnamefont {T.}~\bibnamefont {Calarco}},\ and\ \bibinfo {author} {\bibfnamefont {S.}~\bibnamefont {Montangero}},\ }\href {https://doi.org/10.1088/1361-6633/ac723c} {\bibfield  {journal} {\bibinfo  {journal} {Reports on progress in physics}\ }\textbf {\bibinfo {volume} {85}},\ \bibinfo {pages} {076001} (\bibinfo {year} {2022})}\BibitemShut {NoStop}%
	\bibitem [{\citenamefont {Khaneja}\ \emph {et~al.}(2005)\citenamefont {Khaneja}, \citenamefont {Reiss}, \citenamefont {Kehlet}, \citenamefont {Schulte-Herbr{\"u}ggen},\ and\ \citenamefont {Glaser}}]{khaneja2005optimal}%
	  \BibitemOpen
	  \bibfield  {author} {\bibinfo {author} {\bibfnamefont {N.}~\bibnamefont {Khaneja}}, \bibinfo {author} {\bibfnamefont {T.}~\bibnamefont {Reiss}}, \bibinfo {author} {\bibfnamefont {C.}~\bibnamefont {Kehlet}}, \bibinfo {author} {\bibfnamefont {T.}~\bibnamefont {Schulte-Herbr{\"u}ggen}},\ and\ \bibinfo {author} {\bibfnamefont {S.~J.}\ \bibnamefont {Glaser}},\ }\href {https://doi.org/10.1016/j.jmr.2004.11.004} {\bibfield  {journal} {\bibinfo  {journal} {Journal of magnetic resonance}\ }\textbf {\bibinfo {volume} {172}},\ \bibinfo {pages} {296} (\bibinfo {year} {2005})}\BibitemShut {NoStop}%
	\bibitem [{\citenamefont {Motzoi}\ \emph {et~al.}(2011)\citenamefont {Motzoi}, \citenamefont {Gambetta}, \citenamefont {Merkel},\ and\ \citenamefont {Wilhelm}}]{Motzoi2011optimal}%
	  \BibitemOpen
	  \bibfield  {author} {\bibinfo {author} {\bibfnamefont {F.}~\bibnamefont {Motzoi}}, \bibinfo {author} {\bibfnamefont {J.~M.}\ \bibnamefont {Gambetta}}, \bibinfo {author} {\bibfnamefont {S.~T.}\ \bibnamefont {Merkel}},\ and\ \bibinfo {author} {\bibfnamefont {F.~K.}\ \bibnamefont {Wilhelm}},\ }\href {https://doi.org/10.1103/PhysRevA.84.022307} {\bibfield  {journal} {\bibinfo  {journal} {Phys. Rev. A}\ }\textbf {\bibinfo {volume} {84}},\ \bibinfo {pages} {022307} (\bibinfo {year} {2011})}\BibitemShut {NoStop}%
	\bibitem [{\citenamefont {Lu}\ \emph {et~al.}(2024)\citenamefont {Lu}, \citenamefont {Joshi}, \citenamefont {San~Dinh},\ and\ \citenamefont {Koch}}]{lu2024optimal}%
	  \BibitemOpen
	  \bibfield  {author} {\bibinfo {author} {\bibfnamefont {Y.}~\bibnamefont {Lu}}, \bibinfo {author} {\bibfnamefont {S.}~\bibnamefont {Joshi}}, \bibinfo {author} {\bibfnamefont {V.}~\bibnamefont {San~Dinh}},\ and\ \bibinfo {author} {\bibfnamefont {J.}~\bibnamefont {Koch}},\ }\href {https://doi.org/10.1088/2399-6528/ad22e5} {\bibfield  {journal} {\bibinfo  {journal} {Journal of Physics Communications}\ }\textbf {\bibinfo {volume} {8}},\ \bibinfo {pages} {025002} (\bibinfo {year} {2024})}\BibitemShut {NoStop}%
	\bibitem [{\citenamefont {Bukov}\ \emph {et~al.}(2018)\citenamefont {Bukov}, \citenamefont {Day}, \citenamefont {Sels}, \citenamefont {Weinberg}, \citenamefont {Polkovnikov},\ and\ \citenamefont {Mehta}}]{bukov2018reinforcement}%
	  \BibitemOpen
	  \bibfield  {author} {\bibinfo {author} {\bibfnamefont {M.}~\bibnamefont {Bukov}}, \bibinfo {author} {\bibfnamefont {A.~G.~R.}\ \bibnamefont {Day}}, \bibinfo {author} {\bibfnamefont {D.}~\bibnamefont {Sels}}, \bibinfo {author} {\bibfnamefont {P.}~\bibnamefont {Weinberg}}, \bibinfo {author} {\bibfnamefont {A.}~\bibnamefont {Polkovnikov}},\ and\ \bibinfo {author} {\bibfnamefont {P.}~\bibnamefont {Mehta}},\ }\href {https://doi.org/10.1103/PhysRevX.8.031086} {\bibfield  {journal} {\bibinfo  {journal} {Phys. Rev. X}\ }\textbf {\bibinfo {volume} {8}},\ \bibinfo {pages} {031086} (\bibinfo {year} {2018})}\BibitemShut {NoStop}%
	\bibitem [{\citenamefont {Niu}\ \emph {et~al.}(2019)\citenamefont {Niu}, \citenamefont {Boixo}, \citenamefont {Smelyanskiy},\ and\ \citenamefont {Neven}}]{niu2019universal}%
	  \BibitemOpen
	  \bibfield  {author} {\bibinfo {author} {\bibfnamefont {M.~Y.}\ \bibnamefont {Niu}}, \bibinfo {author} {\bibfnamefont {S.}~\bibnamefont {Boixo}}, \bibinfo {author} {\bibfnamefont {V.~N.}\ \bibnamefont {Smelyanskiy}},\ and\ \bibinfo {author} {\bibfnamefont {H.}~\bibnamefont {Neven}},\ }\href {https://doi.org/10.1038/s41534-019-0141-3} {\bibfield  {journal} {\bibinfo  {journal} {npj Quantum Information}\ }\textbf {\bibinfo {volume} {5}},\ \bibinfo {pages} {33} (\bibinfo {year} {2019})}\BibitemShut {NoStop}%
	\bibitem [{\citenamefont {Zhang}\ \emph {et~al.}(2019)\citenamefont {Zhang}, \citenamefont {Wei}, \citenamefont {Asad}, \citenamefont {Yang},\ and\ \citenamefont {Wang}}]{zhang2019does}%
	  \BibitemOpen
	  \bibfield  {author} {\bibinfo {author} {\bibfnamefont {X.-M.}\ \bibnamefont {Zhang}}, \bibinfo {author} {\bibfnamefont {Z.}~\bibnamefont {Wei}}, \bibinfo {author} {\bibfnamefont {R.}~\bibnamefont {Asad}}, \bibinfo {author} {\bibfnamefont {X.-C.}\ \bibnamefont {Yang}},\ and\ \bibinfo {author} {\bibfnamefont {X.}~\bibnamefont {Wang}},\ }\href {https://doi.org/10.1038/s41534-019-0201-8} {\bibfield  {journal} {\bibinfo  {journal} {npj Quantum Information}\ }\textbf {\bibinfo {volume} {5}},\ \bibinfo {pages} {85} (\bibinfo {year} {2019})}\BibitemShut {NoStop}%
	\bibitem [{\citenamefont {Sivak}\ \emph {et~al.}(2022)\citenamefont {Sivak}, \citenamefont {Eickbusch}, \citenamefont {Liu}, \citenamefont {Royer}, \citenamefont {Tsioutsios},\ and\ \citenamefont {Devoret}}]{sivak2022model}%
	  \BibitemOpen
	  \bibfield  {author} {\bibinfo {author} {\bibfnamefont {V.~V.}\ \bibnamefont {Sivak}}, \bibinfo {author} {\bibfnamefont {A.}~\bibnamefont {Eickbusch}}, \bibinfo {author} {\bibfnamefont {H.}~\bibnamefont {Liu}}, \bibinfo {author} {\bibfnamefont {B.}~\bibnamefont {Royer}}, \bibinfo {author} {\bibfnamefont {I.}~\bibnamefont {Tsioutsios}},\ and\ \bibinfo {author} {\bibfnamefont {M.~H.}\ \bibnamefont {Devoret}},\ }\href {https://doi.org/10.1103/PhysRevX.12.011059} {\bibfield  {journal} {\bibinfo  {journal} {Phys. Rev. X}\ }\textbf {\bibinfo {volume} {12}},\ \bibinfo {pages} {011059} (\bibinfo {year} {2022})}\BibitemShut {NoStop}%
	\bibitem [{\citenamefont {Johansson}\ \emph {et~al.}(2012)\citenamefont {Johansson}, \citenamefont {Nation},\ and\ \citenamefont {Nori}}]{johansson2012qutip}%
	  \BibitemOpen
	  \bibfield  {author} {\bibinfo {author} {\bibfnamefont {J.~R.}\ \bibnamefont {Johansson}}, \bibinfo {author} {\bibfnamefont {P.~D.}\ \bibnamefont {Nation}},\ and\ \bibinfo {author} {\bibfnamefont {F.}~\bibnamefont {Nori}},\ }\href {https://doi.org/10.1016/j.cpc.2012.02.021} {\bibfield  {journal} {\bibinfo  {journal} {Computer physics communications}\ }\textbf {\bibinfo {volume} {183}},\ \bibinfo {pages} {1760} (\bibinfo {year} {2012})}\BibitemShut {NoStop}%
	\bibitem [{\citenamefont {Virtanen}\ \emph {et~al.}(2020)\citenamefont {Virtanen}, \citenamefont {Gommers}, \citenamefont {Oliphant}, \citenamefont {Haberland}, \citenamefont {Reddy}, \citenamefont {Cournapeau}, \citenamefont {Burovski}, \citenamefont {Peterson}, \citenamefont {Weckesser}, \citenamefont {Bright}, \citenamefont {{van der Walt}}, \citenamefont {Brett}, \citenamefont {Wilson}, \citenamefont {Millman}, \citenamefont {Mayorov}, \citenamefont {Nelson}, \citenamefont {Jones}, \citenamefont {Kern}, \citenamefont {Larson}, \citenamefont {Carey}, \citenamefont {Polat}, \citenamefont {Feng}, \citenamefont {Moore}, \citenamefont {{VanderPlas}}, \citenamefont {Laxalde}, \citenamefont {Perktold}, \citenamefont {Cimrman}, \citenamefont {Henriksen}, \citenamefont {Quintero}, \citenamefont {Harris}, \citenamefont {Archibald}, \citenamefont {Ribeiro}, \citenamefont {Pedregosa}, \citenamefont {{van Mulbregt}},\ and\ \citenamefont {{SciPy 1.0 Contributors}}}]{2020SciPy-NMeth}%
	  \BibitemOpen
	  \bibfield  {author} {\bibinfo {author} {\bibfnamefont {P.}~\bibnamefont {Virtanen}}, \bibinfo {author} {\bibfnamefont {R.}~\bibnamefont {Gommers}}, \bibinfo {author} {\bibfnamefont {T.~E.}\ \bibnamefont {Oliphant}}, \bibinfo {author} {\bibfnamefont {M.}~\bibnamefont {Haberland}}, \bibinfo {author} {\bibfnamefont {T.}~\bibnamefont {Reddy}}, \bibinfo {author} {\bibfnamefont {D.}~\bibnamefont {Cournapeau}}, \bibinfo {author} {\bibfnamefont {E.}~\bibnamefont {Burovski}}, \bibinfo {author} {\bibfnamefont {P.}~\bibnamefont {Peterson}}, \bibinfo {author} {\bibfnamefont {W.}~\bibnamefont {Weckesser}}, \bibinfo {author} {\bibfnamefont {J.}~\bibnamefont {Bright}}, \bibinfo {author} {\bibfnamefont {S.~J.}\ \bibnamefont {{van der Walt}}}, \bibinfo {author} {\bibfnamefont {M.}~\bibnamefont {Brett}}, \bibinfo {author} {\bibfnamefont {J.}~\bibnamefont {Wilson}}, \bibinfo {author} {\bibfnamefont {K.~J.}\ \bibnamefont {Millman}}, \bibinfo {author} {\bibfnamefont {N.}~\bibnamefont {Mayorov}}, \bibinfo {author} {\bibfnamefont
	  {A.~R.~J.}\ \bibnamefont {Nelson}}, \bibinfo {author} {\bibfnamefont {E.}~\bibnamefont {Jones}}, \bibinfo {author} {\bibfnamefont {R.}~\bibnamefont {Kern}}, \bibinfo {author} {\bibfnamefont {E.}~\bibnamefont {Larson}}, \bibinfo {author} {\bibfnamefont {C.~J.}\ \bibnamefont {Carey}}, \bibinfo {author} {\bibfnamefont {{\.I}.}~\bibnamefont {Polat}}, \bibinfo {author} {\bibfnamefont {Y.}~\bibnamefont {Feng}}, \bibinfo {author} {\bibfnamefont {E.~W.}\ \bibnamefont {Moore}}, \bibinfo {author} {\bibfnamefont {J.}~\bibnamefont {{VanderPlas}}}, \bibinfo {author} {\bibfnamefont {D.}~\bibnamefont {Laxalde}}, \bibinfo {author} {\bibfnamefont {J.}~\bibnamefont {Perktold}}, \bibinfo {author} {\bibfnamefont {R.}~\bibnamefont {Cimrman}}, \bibinfo {author} {\bibfnamefont {I.}~\bibnamefont {Henriksen}}, \bibinfo {author} {\bibfnamefont {E.~A.}\ \bibnamefont {Quintero}}, \bibinfo {author} {\bibfnamefont {C.~R.}\ \bibnamefont {Harris}}, \bibinfo {author} {\bibfnamefont {A.~M.}\ \bibnamefont {Archibald}}, \bibinfo {author}
	  {\bibfnamefont {A.~H.}\ \bibnamefont {Ribeiro}}, \bibinfo {author} {\bibfnamefont {F.}~\bibnamefont {Pedregosa}}, \bibinfo {author} {\bibfnamefont {P.}~\bibnamefont {{van Mulbregt}}},\ and\ \bibinfo {author} {\bibnamefont {{SciPy 1.0 Contributors}}},\ }\href {https://doi.org/10.1038/s41592-019-0686-2} {\bibfield  {journal} {\bibinfo  {journal} {Nature Methods}\ }\textbf {\bibinfo {volume} {17}},\ \bibinfo {pages} {261} (\bibinfo {year} {2020})}\BibitemShut {NoStop}%
	\bibitem [{\citenamefont {Kadowaki}\ and\ \citenamefont {Nishimori}(1998)}]{Kadowaki1998quantum}%
	  \BibitemOpen
	  \bibfield  {author} {\bibinfo {author} {\bibfnamefont {T.}~\bibnamefont {Kadowaki}}\ and\ \bibinfo {author} {\bibfnamefont {H.}~\bibnamefont {Nishimori}},\ }\href {https://doi.org/10.1103/PhysRevE.58.5355} {\bibfield  {journal} {\bibinfo  {journal} {Phys. Rev. E}\ }\textbf {\bibinfo {volume} {58}},\ \bibinfo {pages} {5355} (\bibinfo {year} {1998})}\BibitemShut {NoStop}%
	\bibitem [{\citenamefont {Lucas}(2014)}]{lucas2014ising}%
	  \BibitemOpen
	  \bibfield  {author} {\bibinfo {author} {\bibfnamefont {A.}~\bibnamefont {Lucas}},\ }\href {https://doi.org/10.3389/fphy.2014.00005} {\bibfield  {journal} {\bibinfo  {journal} {Frontiers in physics}\ }\textbf {\bibinfo {volume} {2}},\ \bibinfo {pages} {5} (\bibinfo {year} {2014})}\BibitemShut {NoStop}%
	\bibitem [{\citenamefont {Eckardt}(2017)}]{eckardt2017colloquium}%
	  \BibitemOpen
	  \bibfield  {author} {\bibinfo {author} {\bibfnamefont {A.}~\bibnamefont {Eckardt}},\ }\href {https://doi.org/10.1103/RevModPhys.89.011004} {\bibfield  {journal} {\bibinfo  {journal} {Rev. Mod. Phys.}\ }\textbf {\bibinfo {volume} {89}},\ \bibinfo {pages} {011004} (\bibinfo {year} {2017})}\BibitemShut {NoStop}%
	\bibitem [{\citenamefont {Bukov}\ \emph {et~al.}(2016)\citenamefont {Bukov}, \citenamefont {Heyl}, \citenamefont {Huse},\ and\ \citenamefont {Polkovnikov}}]{Bukov2016heating}%
	  \BibitemOpen
	  \bibfield  {author} {\bibinfo {author} {\bibfnamefont {M.}~\bibnamefont {Bukov}}, \bibinfo {author} {\bibfnamefont {M.}~\bibnamefont {Heyl}}, \bibinfo {author} {\bibfnamefont {D.~A.}\ \bibnamefont {Huse}},\ and\ \bibinfo {author} {\bibfnamefont {A.}~\bibnamefont {Polkovnikov}},\ }\href {https://doi.org/10.1103/PhysRevB.93.155132} {\bibfield  {journal} {\bibinfo  {journal} {Phys. Rev. B}\ }\textbf {\bibinfo {volume} {93}},\ \bibinfo {pages} {155132} (\bibinfo {year} {2016})}\BibitemShut {NoStop}%
	\bibitem [{\citenamefont {Abanin}\ \emph {et~al.}(2015)\citenamefont {Abanin}, \citenamefont {De~Roeck},\ and\ \citenamefont {Huveneers}}]{Abanin2015exponentially}%
	  \BibitemOpen
	  \bibfield  {author} {\bibinfo {author} {\bibfnamefont {D.~A.}\ \bibnamefont {Abanin}}, \bibinfo {author} {\bibfnamefont {W.}~\bibnamefont {De~Roeck}},\ and\ \bibinfo {author} {\bibfnamefont {F.~m.~c.}\ \bibnamefont {Huveneers}},\ }\href {https://doi.org/10.1103/PhysRevLett.115.256803} {\bibfield  {journal} {\bibinfo  {journal} {Phys. Rev. Lett.}\ }\textbf {\bibinfo {volume} {115}},\ \bibinfo {pages} {256803} (\bibinfo {year} {2015})}\BibitemShut {NoStop}%
	\bibitem [{\citenamefont {Mori}\ \emph {et~al.}(2016)\citenamefont {Mori}, \citenamefont {Kuwahara},\ and\ \citenamefont {Saito}}]{Mori2016rigorous}%
	  \BibitemOpen
	  \bibfield  {author} {\bibinfo {author} {\bibfnamefont {T.}~\bibnamefont {Mori}}, \bibinfo {author} {\bibfnamefont {T.}~\bibnamefont {Kuwahara}},\ and\ \bibinfo {author} {\bibfnamefont {K.}~\bibnamefont {Saito}},\ }\href {https://doi.org/10.1103/PhysRevLett.116.120401} {\bibfield  {journal} {\bibinfo  {journal} {Phys. Rev. Lett.}\ }\textbf {\bibinfo {volume} {116}},\ \bibinfo {pages} {120401} (\bibinfo {year} {2016})}\BibitemShut {NoStop}%
	\bibitem [{\citenamefont {Schindler}\ and\ \citenamefont {Bukov}(2024{\natexlab{a}})}]{Schindler2024counterdiabtic}%
	  \BibitemOpen
	  \bibfield  {author} {\bibinfo {author} {\bibfnamefont {P.~M.}\ \bibnamefont {Schindler}}\ and\ \bibinfo {author} {\bibfnamefont {M.}~\bibnamefont {Bukov}},\ }\href {https://doi.org/10.1103/PhysRevLett.133.123402} {\bibfield  {journal} {\bibinfo  {journal} {Phys. Rev. Lett.}\ }\textbf {\bibinfo {volume} {133}},\ \bibinfo {pages} {123402} (\bibinfo {year} {2024}{\natexlab{a}})}\BibitemShut {NoStop}%
	\bibitem [{\citenamefont {Schindler}\ and\ \citenamefont {Bukov}(2024{\natexlab{b}})}]{schindler2024geometric}%
	  \BibitemOpen
	  \bibfield  {author} {\bibinfo {author} {\bibfnamefont {P.~M.}\ \bibnamefont {Schindler}}\ and\ \bibinfo {author} {\bibfnamefont {M.}~\bibnamefont {Bukov}},\ }\bibfield  {journal} {\bibinfo  {journal} {arXiv preprint arXiv:2410.07029}\ }\href {https://doi.org/10.48550/arXiv.2410.07029} {10.48550/arXiv.2410.07029} (\bibinfo {year} {2024}{\natexlab{b}})\BibitemShut {NoStop}%
	\end{thebibliography}
%

\end{document}